\documentclass[reprint,aps,showpacs]{revtex4}
\usepackage{amsfonts,amssymb,amsmath,epsfig,epic,color}
\usepackage{graphicx}
\usepackage{dcolumn}
\usepackage{bm}

\newcommand{\be}{\begin{equation}}
\newcommand{\ee}{\end{equation}}
\newcommand{\beqs}{\begin{eqnarray}}
\newcommand{\eeqs}{\end{eqnarray}}

\begin{document}
\title{A Many-body Problem with Point Interactions \\ on Two Dimensional Manifolds}

\author{Fatih Erman}
\email{fatih.erman@gmail.com} \affiliation{Department of
Mathematics, Izmir Institute of Technology, Urla,
35430, Izmir, Turkey}%
\author{O. Teoman Turgut}
\email{turgutte@boun.edu.tr} \affiliation{Department of Physics,
Bo\u{g}azi\c{c}i University, Bebek, 34342,
Istanbul, Turkey}%
\altaffiliation[Also at ]{Feza G\"{u}rsey Institute, Kuleli
Mahallesi, \c{S}ekip Ayhan \"{O}z{\i}\c{s}{\i}k Caddesi, No: 44,
Kandilli, 34684, Istanbul, Turkey.}


\begin{abstract} A non-perturbative renormalization of a many-body
problem, where non-relativistic bosons living on a two dimensional
Riemannian manifold interact with each other via the two-body
Dirac delta potential, is given by the help of the heat kernel
defined on the manifold. After this  renormalization procedure,
the resolvent becomes a well-defined operator expressed in terms
of an operator (called principal operator) which includes all the
information about the spectrum. Then, the ground state energy is
found in the mean field approximation and we prove that it grows
exponentially with the number of bosons. The renormalization group
equation (or Callan-Symanzik equation) for the principal operator
of the model  is derived and the $\beta$ function is exactly
calculated for the general case, which includes all particle
numbers.
\end{abstract}


\maketitle

\section{Introduction}

Ultraviolet divergences appear not only in quantum field theories
\cite{peskin} but also in many-body theories and non-relativistic
quantum mechanical problems in which the interaction has a
peculiar singular behavior at short distances \cite{Albeverio
2004, thorn, beg, huang, jackiw}. In all these cases, infinities
are encountered when we calculate some observables (experimentally
measured quantities) e.g., differential cross section of a scattering process,
bound state energy, etc. In order to circumvent these divergences,
a series of algorithmic steps must be applied, and this whole
procedure is called renormalization. The basic idea of
renormalization is first to regularize the infinite integrals by
modifying the short distance (or large momenta) behavior of the
interactions for ultraviolet divergences. This can be accomplished
in several ways with the assumption that the theory is valid up to
a scale determined by an unknown parameter, called cutoff
$\epsilon$ (or $\Lambda$ in momentum space). According to the
modern point of view of renormalization \cite{wilson}, a
renormalizable theory could be regarded as an effective low energy
theory valid up to some unknown energy scale and it is an
approximation to a more fundamental theory beyond this scale.
After having introduced this cutoff parameter $\epsilon$, all the
measured quantities that we are considering in the theory
and the parameters given in the Hamiltonian become
dependent on it. At
this stage, if we remove the cutoff parameter we again encounter
the divergent results for the observables. However,  if we think
one of the parameters in the theory (e.g., coupling constant) as a
function of $\epsilon$ and  relate it to an observable (e.g.,
bound state energy of the system), by solving the appropriate set
of equations, we may remove the dependence on this unknown scale.
That is to say, we can find finite and sensible results for the
other observables in the system (such as differential cross section, phase
shift) by substituting the expression for the coupling constant
found in the previous step and removing $\epsilon$ at the end. If all the
observables are still finite after this awkward procedure, the
theory is said to be renormalizable. If not, one must continue to
apply the same procedure for other remaining parameters (such as
charge, mass, etc.) until every observable becomes finite. This
renormalization procedure can usually be done perturbatively and
only few non-perturbative approaches are available since most
quantum field theories are not exactly solvable.

When de Broglie wavelength of a particle is much larger than the
range of the potential, the interaction can well be approximated
by a Dirac delta function (point interaction). This problem in one
dimension is rather easy and its solution is given in any standard
textbook in quantum mechanics. If we extend this problem into the
one where a particle scatters off a periodic set of delta function
potentials, it is one of the few completely solvable models
\cite{kronig}, which describes the electrons moving in a one
dimensional crystal lattice. In two and three dimensions, the
point interactions give rise to infinities but this problem can be
cured with the renormalization procedure
\cite{huang,jackiw,manuel2}. Most concepts in field theory, such
as dimensional transmutation, regularization, renormalization
group, etc. can be understood in this simpler context. Beside the
role that it plays in understanding renormalization, it has many
applications in diverse areas of physics, as well (see the
references in \cite{Albeverio 2004, Camblong2}).

Point interactions are also considered in a more rigorous context,
so-called self-adjoint extension theory developed by Von Neumann
and a systematic exposition of this subject has been discussed
thoroughly in the monograph \cite{Albeverio 2004}, where a brief
history and an extensive bibliography of it is also given. The
formal Hamiltonian in $D$ dimensions
\be H=-{\hbar^2 \over 2m} \nabla^2 - \lambda
\delta^{(D)}(\textbf{x}) \label{diracdeltaH} \ee
can be rigorously defined as a self-adjoint extension of a free
formally Hermitian Hamiltonian $H_0$ on a space with one point
removed, where the delta center is located and a boundary
condition for the wave function at that point is introduced
\cite{jackiw}. Moreover, there is another rigorous approach to the
above problem where a relation between the resolvents of two
different self-adjoint extensions of one symmetric operator is
given and it is called Krein's formula. The discussion of it for
point interactions has been given in \cite{albeveriokurasov}.
Within this formalism, one can immediately investigate the
spectral properties of the point interactions whereas the domain
issues of the operators can be preferably handled in the Von
Neumann's approach. The results of the self-adjoint extension
methods and the renormalization approach to the point interactions
are the same if a certain relation between the parameter of the
extension and the renormalized (or bare) coupling constant is
satisfied \cite{jackiw}.

Many-body version of the point interactions is also extensively
discussed in the literature from various directions. The
Hamiltonian of the system, in which $n$ particles of mass $m$ are
interacting through the two-body Dirac delta interaction, is
\be H= -{\hbar^2 \over 2m} \sum_{i=1}^{n} \nabla^{2}_{i} - \lambda
\sum_{i<j=1}^{n}\delta^{(D)}(\textbf{x}_i-\textbf{x}_j)\;,
\label{mbodydiracdeltaH} \ee
where $\lambda$ is the coupling constant. One of the earliest
studies on the many-body or few-body version of this model in two
or three dimensions dates back to the work of G. Flamand
\cite{flamand} and the unpublished thesis of J. Hoppe
\cite{Hoppe}, and the ones in the Soviet Union, see the references
given in \cite{Albeverio 2004}. More recently, a perturbative
renormalization to the above $n$-body problem has been worked out
in \cite{adhikari} and also three-body problem in two dimensions
is discussed in \cite{henderson}. It has been proved that the
perturbative treatment of the three-body problem shows new
divergences in three dimensions after the renormalization of the
two-body sector of the problem and these divergences appear for
each added new particles \cite{adhikari}. Therefore, $n-1$ new
scales emerge after the renormalization of the $n$-body problem.
The same model is also rigorously studied in \cite{dellantonio}.

In one dimension, there is no divergence at all and the ground
state of this many-body problem is exactly soluble \cite{mcguire}
and Hartree approximation gives  exactly the same results for
large values of $n$ \cite{calogero}. Moreover, the same problem
for the repulsive case is worked out in  \cite{liebliniger} and
$S$-matrix approach for both the attractive and the repulsive
cases has been studied in \cite{yang1, yang2}

A quantum problem where a single particle interacts with a Dirac
delta potential in two dimensions shows also an elementary example
of dimensional transmutation \cite{huang, thorn, Camblong} (this term is originally introduced in \cite{Coleman}). Under the
scaling transformation $x\rightarrow \alpha x$, the Laplacian and
$\delta^{(2)}(x)$ function transform similarly. In other words,
they have the same dimensions $[L]^{-2}$ so that the coupling constant
$\lambda$ is dimensionless in natural units. Therefore,
Hamiltonian (\ref{diracdeltaH}) in two dimensions does not contain
any intrinsic energy scale due to the dimensionless coupling
constant. A new parameter specifying the bound state energy
is introduced after the renormalization procedure, which then
fixes the energy scale of the system and this phenomenon is called
dimensional transmutation. In fact, as shown in \cite{jackiw}, the time dependent version of this problem has a larger
symmetry group $SO(2,1)$, which exhibits one of the simplest examples of anomaly or quantum mechanical symmetry
breaking. Furthermore, renormalization group (RG)
equations of point interactions have been discussed in
\cite{adhikari, Tarrach1,Camblong2}. The $\beta$ function has been
calculated exactly in there and the theory has been found as
asymptotically free in two dimensions. The RG equations for the two dimensional many-body
extension of the problem, where the Hamiltonian is given by
(\ref{mbodydiracdeltaH}) for $D=2$, has been addressed for
two-body sector in \cite{bergman1, Bergman}. They are
especially useful in this case since there is no analytic solution to the
problem.

S. G. Rajeev \cite{rajeevbound} introduced a new non-perturbative
renormalization method developed for bound state problems of some
quantum many-body theories: fermionic and bosonic quantum fields
interacting with a point source with two internal states and
non-relativistic bosons interacting via two-body point
interactions. One of the main advantages of this approach is that
all the information about the spectrum of the model is described
by an explicit formula instead of imposing the boundary conditions
on the operator as in the case of self-adjoint extension theory.
Another advantage is that the renormalization is performed
non-perturbatively by introducing fictitious degrees of freedom
via othofermion algebra so that it helps us to reduce the renormalization to
simply normal ordering of an operator which is called principal
operator $\Phi$ and then all the information about the spectrum of
the problem can be found from its explicit well-defined form. Due
to the  non-perturbative nature of this method it is also
particularly useful for dealing with the bound state problems. We
are not going to review the original method developed in there.
Instead, we suggest the reader to read through the relevant parts
of the paper \cite{rajeevbound}, especially $\lambda
\phi^{4}_{(2+1)NR}$ model  to make the reading of this paper
easier (the problem where bosons interacts with each other via
two-body Dirac delta potentials is indeed known as the formal
non-relativistic limit of the $\lambda \phi^4$ scalar field theory
\cite{jackiw, begfurlong, dimock1}). A mathematically more
rigorous discussion for this approach to $\lambda
\phi^{4}_{(2+1)NR}$ has been given in \cite{rajeevdimock}.

Following the original ideas developed in \cite{rajeevbound}, we
previously considered \textit{the bound state problem} for $N$
point interactions in two and three dimensional Riemannian
manifolds \cite{ermanturgut} by using the heat kernel and
discussed its spectral properties in there. The same model from
the Krein's point of view has been discussed for special explicit
manifolds, such as strips or tubes \cite{exnerstrip,exnertube},
and it is considered as a natural model for quantum wires
including point like impurities. The model that we will now
construct is the many-body version of our previous work
\cite{ermanturgut}, where the non-relativistic bosons interact
with each other via two-body Dirac delta function potential. Our
primary motivation here is to find a better understanding of the
renormalization of many-body models on Riemannian manifolds.

The paper is organized as follows. In Section II, we construct a
model where the non-relativistic bosons interact with each other
via two-body Dirac delta potential in two dimensional Riemannian
manifolds. This construction is motivated by the work
\cite{rajeevbound} where a new non-perturbative renormalization
method is developed. By extending the Fock space, it becomes
possible to renormalize the model non-perturbatively by simply
normal ordering of an operator, called principal operator. Section
III is about the mean field approximation of the model and it has
been found that the magnitude of the ground state energy grows
exponentially with the number of bosons, which agrees with answer
in the flat case already found in \cite{rajeevbound}. The same
formulation can also be applied to the one dimensional model where
there is no renormalization. In this case, the mean field
approximation that we develop here gives exactly the same result
with the one given in the literature \cite{mcguire, calogero}.
Finally, we proceed with the renormalization group equations for
this model and the $\beta$ function is exactly calculated.

\section{Construction of Model}
\label{construction}

The non-perturbative method which is applied to our system here consists of the following series of steps:
\begin{enumerate}
  \item[1)] We first regularize the Hamiltonian via heat kernel
  \item[2)] Then we extend the Fock space by the help of orthofermion algebra so that new Fock space becomes a direct sum of
  two Hilbert spaces.
  \item[3)] The Hamiltonian on the extended Fock space is constructed in such a way that the regularized resolvent projected onto the old Fock space
  gives an equivalent expression for the regularized resolvent of our original Hamiltonian. Hence, the coupling constant becomes additive rather than multiplicative.
  \item[4)] By normal ordering the equivalent expression of the regularized resolvent, the singular part of the problem becomes transparent due to the short time asymptotic expansion of the heat kernel. Then, it is possible to choose the coupling constant in such a way that the singular part is removed.
\end{enumerate}

The Hamiltonian on a two dimensional Riemannian manifold
$(\mathcal{M}$, g) is formally given in the second quantized
language (we use the units such that $\hbar = 2m =1$)
\begin{widetext}
\beqs
H = - \int_{\mathcal{M}} \mathrm{d}_{g}^{2} x \;
\phi^{\dag}_{g}(x)\, \nabla_{g}^2 \, \phi_g(x) - {\lambda \over 2}
\int_{\mathcal{M}^2} \mathrm{d}_{g}^{2} x \, \mathrm{d}_{g}^{2} x'
\; \phi^{\dag}_{g}(x') \, \phi^{\dag}_{g}(x) \,
\delta_g^{(2)}(x,x') \, \phi_{g}(x) \, \phi_{g}(x') \;,
\label{ham}
\eeqs
\end{widetext}
where $\mathrm{d}_{g}^{2} x = \sqrt{\det g} \; \mathrm{d}x^1
\mathrm{d}x^2$ is the two dimensional volume element,
$\nabla_{g}^2$ is the Laplace-Beltrami operator (or simply
Laplacian) defined in a local coordinate system, also written as
$x \equiv(x^{1}, x^{2})$
\be
\nabla^2_g = \frac{1}{\sqrt{\mathrm{det}\,g}} \sum_{i,j=1}^{2}
\frac{\partial}{\partial x^i} \left(g^{ij} \, \sqrt{\mathrm{det}
\,g} \; \frac{\partial}{\partial x^j}\right)\;,
\ee
and $\lambda$ is a positive coupling constant (it corresponds to
an attractive interaction). Here, $\phi^{\dag}_{g}(x)$,
$\phi_{g}(x)$ are the bosonic creation-annihilation operators and
$\delta_g^{(2)}(x,x')$ is the Dirac delta function defined on the
two dimensional Riemannian manifold with metric structure $g$:
\be
\int_{\mathcal{M}} \mathrm{d}^{2}_{g}x \;\delta_g^{(2)}(x,x')
f(x')= f(x) \;.
\ee
It is important to notice that the number of bosons
$\int_{\mathcal{M}} \mathrm{d}_g^{2} x
\;\phi^{\dag}_{g}(x)\,\phi_{g}(x)$ is conserved in our model.

Let us suppose that there exists a negative bound state energy
$E_b < 0$ corresponding to the normalized wave function
$\psi(x_1,\ldots,x_n;g)$, that is,
\be
\int_{\mathcal{M}^{n}} \mathrm{d}^{2}_{g} x_1 \ldots
\mathrm{d}^{2}_{g} x_n \; |\psi(x_1,\ldots,x_n;g)|^2 =1 \;.
\ee
Due to scale invariance of the Hamiltonian under the
transformation $g \rightarrow \alpha^2 g$ with a positive constant
$\alpha^2$, the wave function $\psi(x_1,\ldots,x_n;g)=\alpha^n
\psi(x_1,\ldots,x_n;\alpha^2 g)$ satisfies the same eigenvalue
equation with the energy $-\alpha^2 |E_b|$. Therefore, the
existence of a negative bound state energy implies that the energy
can be made arbitrarily negative by choosing arbitrarily large
values of $\alpha$. This means that the energy is not bounded from
below, which is not allowed in a sensible theory.

In order to cure the problem, we will first regularize the model.
The same model in flat space has been discussed in
\cite{rajeevbound,rajeevdimock} and the renormalization has been
performed in a non-perturbative way. In that case, the divergence
appears due to the large values of momenta (ultraviolet), or short
distances. Hence, we expect that the ultraviolet divergence must
also exist for the same model defined on manifold since every
Riemannian manifold can locally be considered as a flat space. In
\cite{ermanturgut}, we have proved that the divergence due to
short distance is replaced with the short ``time" for a simplified
version of this model, where a particle interacts with several
external delta potentials on a manifold. This is accomplished by
expressing the resolvent of the system in terms of the heat
kernel. In this way, we have been able to subtract the divergence
from our model by using the short ``time" asymptotic behavior of
the heat kernel. This motivates us that the proper regularization
for the many-body version must also be performed via heat kernel
and a natural choice for the regularized Hamiltonian is
\begin{widetext}
\beqs
&\ & \hskip-1cm H^{\epsilon} = H_0
 - {\lambda(\epsilon) \over 2}\int_{\mathcal{M}^5}
\mathrm{d}_g^2 x_1 \mathrm{d}_g^2 x'_1 \mathrm{d}_g^2 x_2
\mathrm{d}_g^2 x_2' \mathrm{d}_g^2 y \; \phi^{\dag}_{g}(x_1)
\phi^{\dag}_{g}(x_2) K_{\epsilon} (x_1,y;g) K_{\epsilon} (y,x_2;g)
\cr &\ & \hspace{4cm} \times \; K_{\epsilon} (x_1',y;g)
K_{\epsilon} (y,x'_2;g) \phi_{g}(x_1')\phi_{g}(x_2') \;,
\label{regularized H}
\eeqs
\end{widetext}
with $\epsilon$ the short ``time" cutoff parameter, $H_0$ the
free Hamiltonian, and $K_{\epsilon} (x,y;g)$ the heat kernel on
the manifold defined as a fundamental solution to the heat
equation \cite{chavel2}
\be
{\partial K_t(x,y;g) \over
\partial t} = \nabla_{g}^{2}K_t(x,y;g) \;. \label{heatequation}
\ee
Unless otherwise stated, it is always assumed that the Laplacian
$\nabla_{g}^{2}$ acts on the functions of the variable $x$. One of
the most important properties of the heat kernel that we use in
this paper is that it converges to Dirac delta function
\be
K_t(x,y;g) \rightarrow \delta_{g}^{(2)}(x,y) \;,
\label{heatkernel delta}
\ee
as $t \rightarrow 0^+$ in the sense of distributions. It is also
symmetric $K_t(x,y;g)=K_t(y,x;g)$ for all $x,y\in \mathcal{M}$ and $t>0$ \cite{chavel2}. If we remove the
cutoff, that is, take the limit $\epsilon \rightarrow 0^+$, we
immediately see that we recover the original Hamiltonian given in
(\ref{ham}). It should also be pointed out that we consider the
coupling constant in (\ref{regularized H}) as a function of the
cutoff $\epsilon$, and its explicit form will be determined
later.

Now, we will consider the resolvent of the Hamiltonian (\ref{ham})
in Fock space $\mathcal{F_B}$ with arbitrary number of bosons.
Following the method developed for the same model in the
plane \cite{rajeevbound}, we will extend the bosonic Fock space
$\mathcal{F_B}$ that we have started with to a larger Fock space,
as it was first introduced in \cite{rajeevbound}. For this purpose, we define new creation and annihilation operators, which
obey orthofermionic algebra \cite{mishra}:
\beqs
\chi_{g}(x)\chi^{\dag}_{g}(y) &=&  \delta_{g}^{(2)}(x,y)
\Pi_0, \cr \chi_{g}(x)\chi_{g}(y)&=&0=
\chi^{\dag}_{g}(x)\chi^{\dag}_{g}(y) \;, \label{orthofermion
algebra}
\eeqs
where
\be
\Pi_1 = \int_{\mathcal{M}} \mathrm{d}_{g}^2 x \,
\chi^{\dag}_{g}(x) \chi_{g}(x) , \;\;\;\Pi_0 = 1- \Pi_1 \;
\ee
are the projection operators onto one-orthofermion and no-orthofermion states,
respectively. It follows easily that there can be at most one
orthofermion in any state. The new Fock space is introduced as a direct sum of two Hilbert spaces
\be
\mathcal{\tilde{F}_B} = \mathcal{F_B} \oplus \bigg[ \mathcal{F_B}
\otimes \mathcal{L}^2(\mathcal{M}) \bigg] \;,
\ee
where the first sector which does not include any orthofermion is written as bosonic Fock space $\mathcal{F_B}$ and the second sector with a single
orthofermion as $\mathcal{F_B}
\otimes \mathcal{L}^2(\mathcal{M})$. Here, we identify
the space of single orthofermion states by $\mathcal{L}^2(\mathcal{M})$.

The advantage of introducing this trick is
that it allows us to rewrite the resolvent of the model in such a
manner that the coupling constant appears additively rather than
multiplicatively. Actually, the idea of introducing unphysical
particles in such a way as to cancel the infinities is not a new
idea (see the references in \cite{schweber}). As a result, we will
be able to subtract the divergence from our model
nonperturbatively by simply normal ordering the operators. Now we
define the augmented regularized Hamiltonian
$\tilde{H}^{\epsilon}$ on $\mathcal{\tilde{F_B}}$ as
\begin{widetext}
\beqs
& &
 \tilde{H}^{\epsilon} = H_0 \Pi_0 + \bigg[{1 \over \sqrt{2}} \int_{\mathcal{M}^3} \mathrm{d}_g^2 x_1 \mathrm{d}_g^2 x_2
\mathrm{d}_g^2 y \; \phi^{\dag}_{g}(x_1)\, \phi^{\dag}_{g}(x_2)
K_{\epsilon}(x_1,y;g) K_{\epsilon}(y, x_2;g)\chi_{g}(y) + h.c.
\bigg] + {\Pi_1 \over \lambda(\epsilon)} \;.
\eeqs
\end{widetext}

If we split the Hilbert space according to the orthofermion number, the
corresponding splitting of the operator $\tilde{H}^{\epsilon} - E
\Pi_0$ can be written in the following matrix form
\be
\tilde{H}^{\epsilon} - E \Pi_0 = \left(%
\begin{array}{cc}
  a & b_{\epsilon}^\dagger \\
  b_{\epsilon} & d_{\epsilon} \\
\end{array}%
\right)\;,
\ee
with $a: \mathcal{F_B}\rightarrow \mathcal{F_B}$,
$b^{\dagger}_{\epsilon} : \mathcal{F_B} \otimes
\mathcal{L}^2(\mathcal{M}) \rightarrow \mathcal{F_B}$,
$d_{\epsilon}: \mathcal{F_B} \otimes \mathcal{L}^2(\mathcal{M})
\rightarrow \mathcal{F_B} \otimes \mathcal{L}^2(\mathcal{M}) $.
Accordingly, the explicit form of the matrix elements of the above
matrix is
\begin{widetext}
\beqs
a & = & (H_0 - E)\Pi_0 \;, \hspace{2cm} d_\epsilon= {\Pi_1
\over \lambda(\epsilon)} \cr b^{\dag}_{\epsilon}&=& {1 \over
\sqrt{2}} \int_{\mathcal{M}^3} \mathrm{d}_g^2 x_1 \mathrm{d}_g^2
x_2 \mathrm{d}_g^2 y \; \phi^{\dag}_{g}(x_1)\,
\phi^{\dag}_{g}(x_2) K_{\epsilon}(x_1,y;g)K_{\epsilon}(y, x_2;g)
\chi_{g}(y) \;.
\eeqs
\end{widetext}
Then, one can construct the augmented regularized resolvent
defined as $(\tilde{H}^{\epsilon} -E \Pi_0)^{-1}$ and let us
suppose that it is of the following matrix form
\be
\tilde{R^{\epsilon}}(E)= \left(%
\begin{array}{cc}
  \alpha_{\epsilon} & \beta^\dagger_{\epsilon} \\
  \beta_{\epsilon} & \delta_{\epsilon} \\
\end{array}%
\right)\;.
\ee
Incidentally, the energy $E$ here should be considered as a
complex variable. One can find $\alpha_{\epsilon},
\beta_{\epsilon}, \delta_{\epsilon}$ in terms of $a,
b_{\epsilon}$, and $d_{\epsilon}$ by a direct computation. This
could be done in two apparently different but equivalent  ways and
the formulas were explicitly given in the appendix of
\cite{rajeevbound}. One of the solutions to $\alpha_{\epsilon}$ is
\be
\alpha_{\epsilon} = \left[a- b^\dag_{\epsilon}
\,d^{-1}_{\epsilon} \, b_{\epsilon} \right]^{-1}={1 \over
H^{\epsilon} -E} = R^{\epsilon}(E) \;.
\ee
This means that $\tilde{R}_{\epsilon}(E)$ projected to
$\mathcal{F_B}$ is just the resolvent of the operator
$H^{\epsilon}$. The other solution for $\alpha_{\epsilon}$
\cite{rajeevbound} is
\be
\alpha_{\epsilon}  = a^{-1} +  a^{-1} \, b^\dag_{\epsilon}
\left[d_{\epsilon} - b_{\epsilon}\, a^{-1}\, b^\dag_{\epsilon}
\right]^{-1} b_{\epsilon} \, a^{-1} \;.
\ee
Combining both solutions give
\be
R^{\epsilon}(E) = \alpha_{\epsilon} = a^{-1} +  a^{-1} \,
b^\dag_{\epsilon} \left[\Phi^{\epsilon}(E)\right]^{-1}
b_{\epsilon} \, a^{-1} \;,
\ee
where we have defined
\begin{widetext}
\beqs
 &\ & \Phi^{\epsilon}(E)= {\Pi_1 \over \lambda(\epsilon)}-
 {1 \over 2} \int_{\mathcal{M}^6} \mathrm{d}_g^2 x_1 \mathrm{d}_g^2 x_2
\,\mathrm{d}_g^2 y \, \mathrm{d}_g^2 x_1' \, \mathrm{d}_g^2 x_2'\,
\mathrm{d}_g^2 y' \; K_{\epsilon}(x_1,y;g) K_{\epsilon}(y,x_2;g)
\cr &\ & \hspace{3cm} \times \; K_{\epsilon} (x'_1,y';g)
K_{\epsilon}(y', x_2';g) \chi^{\dag}_{g}(y) \bigg[ \phi_{g}(x_1)
\phi_{g}(x_2) {1\over H_0-E} \phi^{\dagger}_{g}(x'_1)\,
\phi^{\dagger}_{g}(x_2')\bigg] \chi_{g}(y') \;,\label{Phieps}
\eeqs
\end{widetext}
which is called the regularized principal operator, in which the
coupling constant is written additively. Now, in order to see and
separate out the divergent part from (\ref{Phieps}), we will
normal order the operators in (\ref{Phieps}) by using the
commutation relations of the field operators. For simplicity, we
explicitly perform our calculations for compact manifolds here,
but our result is also valid, in principle, for non-compact manifolds by using a
similar method that we have done for non-relativistic Lee model
\cite{nrleemodelonmanifold, ermanturgutlee2}.

In analogy with the
plane wave mode expansion of the field operators in quantum field
theory, one can write the eigenfunction expansion of the creation
and annihilation operators as
\beqs
\phi_{g}^{\dagger}(x)&=& \sum_{l} \phi_{l}^{\dagger} \;
f_{l}(x;g) \cr \phi_g(x)&=& \sum_{l} \phi_l \; f_l(x;g) \;,
\label{eigenexpfield}
\eeqs
where $f_l(x;g)$ is the complete and orthonormal eigenfunction of
the Laplace-Beltrami operator \cite{Rosenberg}:
\beqs
-\nabla_{g}^{2}f_l(x;g)&=& \sigma_l f_l(x;g) \cr \int_{\mathcal{M}} \mathrm{d}_g^2 x \; f_l(x;g) \, f_{m}^{*}(x;g)
& = & \delta_{lm} \cr \sum_l f_l(x;g) \, f_{l}^{*}(y;g) &=& \delta_{g}^{(2)}(x,y)
\label{heat kernel} \;,
\eeqs
with the spectrum $\{0=\sigma_0 \leq \sigma_1 \leq \sigma_2 \leq \ldots \}$ so that the free Hamiltonian becomes
\be
H_0= \sum_l \sigma_l \; \phi^{\dag}_{l}\phi_{l} \;.
\ee
It must be emphasized
that the degeneracy is formally taken into account in the above
sum by the index $l$. For simplicity, we have suppressed this
possible degeneracy. Using the commutation relation $[\phi_l,\phi^{\dag}_{l'}]=\delta_{ll'}$, it is easy to see that
$(H_0-E)\phi^{\dag}_{l}=\phi^{\dag}_{l}(H_0-E+\sigma_l)$.
Multiplying this equation by $(H_0-E)^{-1}$ from left and by $(H_0-E+\sigma_l)^{-1}$ from right we get
\be
(H_0-E)^{-1}\phi^{\dag}_{l}=\phi^{\dag}_{l} (H_0-E+\sigma_l)^{-1} \label{h0inverse} \;.
\ee
We now multiply both sides of the above equation with $f_l(x;g)$ and take the sum over $l$ to obtain
\be
(H_0-E)^{-1}\phi^{\dag}_{g}(x)= \sum_l \phi^{\dag}_{l} \int_{0}^{\infty} \mathrm{d} t \; e^{-t(H_0-E+\sigma_l)} f_l(x;g) \;,
\ee
where we have used the fact $(H_0-E+\sigma_l)^{-1}=\int_{0}^{\infty} \mathrm{d} t \; e^{-t(H_0-E+\sigma_l)}$.
Since $\phi^{\dag}_{l}= \int_{\mathcal{M}} \mathrm{d}_{g}^{2}y \; \phi^{\dag}_{g}(y)
f_{l}^{*}(y;g)$
and the eigenfunction expansion of the heat
kernel is given by \cite{Rosenberg}
\beqs
K_t(x,y;g)= \sum_{l} e^{-t \sigma_l} f_l(x;g)f_{l}^{*}(y;g) \;,
\label{heatkernelexpansion}
\eeqs
we find
\be
(H_0-E)^{-1}\phi^{\dag}_{g}(x)=  \int_{\mathcal{M}} \mathrm{d}_{g}^{2}y \; \phi^{\dag}_{g}(y)
\int_{0}^{\infty} \mathrm{d} t \; e^{-t(H_0-E)} K_t(x,y;g) \;.
\ee
Similarly, by using the same procedure, we
can shift all the creation operators
$\phi^{\dagger}_{g}(x_{1}')\phi^{\dagger}_{g}(x_{2}')$ to the left
\begin{widetext}
\beqs
{1\over H_0-E} \phi^{\dagger}_{g}(x_{1}')\phi^{\dagger}_{g}(x_{2}')
= \int_{\mathcal{M}^2} \mathrm{d}_{g}^{2}y_{1}' \,
\mathrm{d}_{g}^{2}y_{2}'
\;\phi^{\dagger}_{g}(y_{1}')\phi^{\dagger}_{g}(y_{2}') \int_0^\infty
\mathrm{d} t \; e^{-t(H_0-E)} \, K_{t}(x_{1}',y_{1}';g)
K_{t}(x_{2}',y_{2}';g) \;,
\eeqs
\end{widetext}
and then normal order the new expression with the annihilation
operators $\phi_{g}(x_{1})\phi_{g}(x_{2})$ so that we obtain the normal
ordered regularized principal operator
\begin{widetext}
\beqs
&\ & \Phi^{\epsilon}(E)= {\Pi_1 \over \lambda(\epsilon) } -
{1 \over 2} \int_{\mathcal{M}^2} \mathrm{d}_g^2 x \,
\mathrm{d}_g^2 x' \; \chi^{\dag}_{g}(x) \bigg[
\int_{\mathcal{M}^4} \mathrm{d}_g^2 x_1 \, \mathrm{d}_g^2 x_2 \,
\mathrm{d}_g^2 x'_1 \, \mathrm{d}_g^2 x'_2 \;
\phi^{\dag}_{g}(x_1')\,
 \phi^{\dag}_{g}(x_2') \cr &\ & \hspace{2cm} \times \; \int_{0}^{\infty}
 \mathrm{d} t\; K_{t+\epsilon}(x'_1,x';g) \,
 K_{t+\epsilon}(x',x'_2;g) \, K_{t+\epsilon}(x_1,x;g) \,
 K_{t+\epsilon}(x,x_2;g) e^{-t (H_0 -E)}
\phi_{g}(x_1)\, \phi_{g}(x_2) \cr & & \hspace{1cm} + \; 4
\int_{\mathcal{M}^2} \mathrm{d}_g^2 x_1 \, \mathrm{d}_g^2 x_2 \;
\phi^{\dag}_{g}(x_1) \int_{0}^{\infty} \mathrm{d} t \;
K_{t+\epsilon}(x_1,x';g) \, K_{t+2\epsilon}(x',x;g) \,
K_{t+\epsilon}(x, x_2;g)\,
 \, e^{-t(H_0 -E)} \phi_{g}(x_2) \cr &\ & \hspace{7cm}
+ \; 2 \int_{0}^{\infty} \mathrm{d} t \;
K_{t+2\epsilon}^{2}(x,x';g) \, e^{-t(H_0 -E)}
 \bigg] \chi_{g}(x') \;,\label{divergentPhi}
\eeqs
\end{widetext}
where the semi-group property of the heat kernel
\beqs
\label{semigroupprop} \hskip-0.5cm K_{t_1+t_2}(x,y;g)=
\int_{\mathcal{M}} \mathrm{d}_g^2 z \; K_{t_1}(x,z;g)
K_{t_2}(z,y;g)
\eeqs
is used. We expect that as $\epsilon \rightarrow 0^+$ the last
``time" integral in (\ref{divergentPhi}) is divergent since it is
the term that corresponds to the infinite expression in the
principal operator for the flat space $\mathbb{R}^2$, where it has
been discussed in \cite{rajeevbound}. In fact, we can also naively
show that the divergence which appears in the principal operator
(\ref{divergentPhi}) is due to the short ``time" asymptotic
behavior of the heat kernel.

In order to see this, let us find an
upper bound to the expectation value of the last term in the
principal operator (\ref{divergentPhi}) after taking the limit
$\epsilon \rightarrow 0^{+}$. For $(n-2)$-bosonic and one-orthofermion
states
\beqs
| \Psi \rangle = |\psi_{b}^{(n-2)} \rangle \otimes
\int_{\mathcal{M}} \mathrm{d}_g^2 x \, \chi^{\dagger}_{g}(x)
\psi(x) | 0 \rangle \;,
\eeqs
we get for the expectation value
\begin{widetext}
\beqs
& & \langle \Psi| \int_{\mathcal{M}^2} \mathrm{d}_g^2 x
\,\mathrm{d}_g^2 x' \; \chi^{\dag}_{g}(x') \int_{0}^{\infty}
\mathrm{d} t \; K_{t}^{2}(x,x';g)
 e^{-t(H_0 -E)} \chi_{g}(x) | \Psi \rangle \cr & & \hspace{2cm}=
\int_{0}^{\infty} \mathrm{d} t \; \langle \psi_{b}^{(n-2)} |
e^{-t(H_0 -E)} | \psi_{b}^{(n-2)} \rangle \int_{\mathcal{M}^2}
\mathrm{d}_g^2 x \, \mathrm{d}_g^2 x' \;
  \psi^{*}(x')K_{t}(x,x';g) \; K_{t}(x,x';g) \psi(x)
\cr & & \hspace{3cm} \leq \int_{0}^{\infty} \mathrm{d} t \;
\langle \psi_{b}^{(n-2)} | e^{-t(H_0 -E)} | \psi_{b}^{(n-2)}
\rangle \int_{\mathcal{M}^2} \mathrm{d}_g^2 x \, \mathrm{d}_g^2 x'
\; K_{t}^{2}(x,x';g) |\psi(x)|^2 \cr & & \hspace{3cm} \leq
\int_{\mathcal{M}} \mathrm{d}_g^2 x \; \int_{0}^{\infty}
\mathrm{d} t \; \langle \psi_{b}^{(n-2)} | e^{-t(H_0 -E)} |
\psi_{b}^{(n-2)} \rangle
 K_{2t} (x,x;g) |\psi(x)|^2
\cr & & \hspace{3cm} = \int_{\mathcal{M}} \mathrm{d}_g^2 x \;
|\psi(x)|^2 \langle \psi_{b}^{(n-2)} | \int_{0}^{\infty}
\mathrm{d} t \; e^{-t(H_0 -E)} K_{2t} (x,x;g) | \psi_{b}^{(n-2)}
\rangle \;, \label{divergentterm}
\eeqs
\end{widetext}
where we have used the Cauchy-Schwarz inequality with the
semi-group (\ref{semigroupprop}) and symmetry properties of the
heat kernel. Therefore, ``time" integral in the right hand side of
(\ref{divergentterm}) is divergent due to the first term in the
short ``time" asymptotic expansion of the diagonal heat kernel,
which is given by
\be
K_t(x,x;g) \sim {1 \over (4 \pi t)^{D/2}} \sum_{k=0}^{\infty}
u_k(x,x)\; t^{k} \;, \label{asymheat}
\ee
for any $D$ dimensional Riemannian manifold without boundary \cite{gilkey}.
Here $u_k(x,x)$ are scalar polynomials in curvature tensor of the
manifold and its covariant derivatives at point $x \in
\mathcal{M}$. This means that if the left hand side of
(\ref{divergentterm}) is divergent, \textit{this is basically due
to the singular behavior of the heat kernel near $t=0$ in the last
term of the principal operator (\ref{divergentterm})}.

All these
suggest us to choose the bare coupling constant as
\beqs
\label{coupling constant} {1 \over \lambda(\epsilon)} =
\int_{\epsilon}^{\infty} \mathrm{d} t \; {e^{-t \mu^2 } \over 8\pi
t} \;,
\eeqs
where $-\mu^2$ is to be related to (the experimentally determined)
bound state energy of two-boson system. The parameter $\mu^2$ is at present an
arbitrary renormalization scale, which breaks the scale invariance in the unrenormalized problem. Even if there is
no bound state in the spectrum, our prescription will lead to a finite formulation.
Yet, later on we will prove that for sufficiently large values of $\mu^2$ we can always find a two-body bound state and
hence we may solve $\mu^2$ in terms of the physical two-body bound state
energy (see equation (\ref{twobodyEgrmu})). In Section \ref{Renormalization Group},
a different prescription will be used where the renormalization scale
is not directly related to the bound state energy.

With the present choice of the
coupling constant (\ref{coupling constant}), we take the limit $\epsilon\rightarrow 0^+$ in
(\ref{divergentPhi}), and readily obtain
\begin{widetext}
\beqs
&\ & \Phi(E)= \int_{\mathcal{M}^2} \mathrm{d}_g^2 x \,
\mathrm{d}_g^2 x' \chi^{\dag}_{g}(x)
 \int_{0}^{\infty} \mathrm{d} t \; \bigg[ {e^{- t \mu^2 }
 \over 8 \pi t}
\delta_g^{(2)}(x,x')  - K_{t}^{2}(x,x';g) e^{-t(H_0 -E)} \bigg]
\chi_{g}(x') \cr &\ & - {1 \over 2} \int_{\mathcal{M}^2}
\mathrm{d}_g^2 x \,\mathrm{d}_g^2 x' \; \chi^{\dag}_{g}(x)
 \bigg[ \int_{\mathcal{M}^4} \mathrm{d}_g^2 x_1 \, \mathrm{d}_g^2 x_2 \,
\mathrm{d}_g^2 x'_1 \, \mathrm{d}_g^2 x'_2 \;
\phi^{\dag}_{g}(x'_1)\,
 \phi^{\dag}_{g}(x'_2) \int_{0}^{\infty} \mathrm{d} t\; K_{t}(x'_1,x';g)
 \cr &\ & \hskip-1cm \times  \;  K_{t}(x',x'_2;g) \, K_{t}(x_1,x;g) \, K_{t}(x,x_2;g) \,
\, e^{-t (H_0 -E)} \phi_{g}(x_1)\, \phi_{g}(x_2) + 4
\int_{\mathcal{M}^2} \mathrm{d}_g^2 x_1 \, \mathrm{d}_g^2 x_2 \;
\phi^{\dag}_{g}(x_1) \ \cr &\ & \times \int_{0}^{\infty}
\mathrm{d} t \; K_{t}(x_1,x';g) \,K_{t}(x',x;g) \, K_{t}(x,x_2;g)
\, e^{-t (H_0 -E)} \phi_{g}(x_2)
 \bigg] \chi_{g}(x') \;. \label{Phi}
\eeqs
\end{widetext}
This is a well-defined form of the principal operator and we can
show that the choice for the coupling constant (\ref{coupling
constant}) is sufficient to remove the divergence from our
problem. Once we have a proper and well-defined expression of the
principal operator, we expect that all the divergences are removed
since the resolvent which determines the spectrum of the problem
is expressed in terms of it. It must be emphasized here that the
principal operator can be extended to its largest domain of
definition in the complex energy plane by analytic continuation.

We must first note that the behavior of the off-diagonal term of
the heat kernel near $t=0$ is intimately related to the small
distance behavior due to the initial condition given for the heat
kernel. In fact one can show that the choice for the coupling
constant (\ref{coupling constant}) is the appropriate one to get
rid off the infinity by writing the square of the heat kernel in
the following subtle way near $t=0$:
\begin{widetext}
\beqs
&\ & \Phi(E)= \int_{\mathcal{M}^2} \mathrm{d}_g^2 x \,
\mathrm{d}_g^2 x' \chi^{\dag}_{g} (x)
 \int_{0}^{\infty} \mathrm{d} t \; \bigg[ {e^{-t \mu^2 }
 \over 8 \pi t}
\delta_g^{(2)}(x,x')- K_{2t}(x,x';g) \delta_g^{(2)}(x,x')
e^{-t(H_0 -E)} \bigg] \chi_{g}(x')
 \cr & & \hspace{7cm} + \; \mathrm{Regular} \; \mathrm{terms} \;. \label{19}
\eeqs
\end{widetext}
The following heuristic argument can be given to justify this
choice. Here, what we mean by ``regular terms" are the other terms
in (\ref{Phi}) and the ignored terms that is coming from the
outside of the region $t=0$. Let us first look at the matrix
element of the second term in the first ``time" integral in the
principal operator (\ref{Phi}):
\be
 \int_{\mathcal{M}} \mathrm{d}_g^2 x \; \psi_{a}^*(x) \, K_{t}(x,y;g)\,
K_{t}(x,y;g) \, \psi_{b}(y) \;, \label{K^2}
\ee
as $t \rightarrow 0^+$. As a consequence of (\ref{heatkernel
delta}), it is possible to replace the function $\psi_a^*(x)$ by
$\psi_a^*(y)$ in this limit, so that we have
\beqs
&\ & \int_{\mathcal{M}} \mathrm{d}_g^2 x \; \psi_a^*(x) \,
K_{t}(x,y;g)
  \, K_{t}(x,y;g) \, \psi_b(y) \cr &\
& \approx
\psi_a^*(y) \int_{\mathcal{M}} \mathrm{d}_g^2 x \;  K_{t}(x,y;g) \,
 K_{t}(x,y;g) \, \psi_b(y) \cr
 &\ & \approx \psi_a^*(y)\, K_{2t}(y,y;g) \, \psi_b(y) \;,
\eeqs
where we have used the semi-group property of the heat kernel
(\ref{semigroupprop}). Therefore, if we take the integral
(\ref{19}) over $x'$ and substitute the first term in the
asymptotic expansion (\ref{asymheat}) of the diagonal heat kernel
as $t \rightarrow 0^+$, we get
\begin{widetext}
\beqs
&\ & \hskip-1cm \Phi(E)= \int_{\mathcal{M}} \mathrm{d}_g^2 x
\; \chi^{\dag}_{g}(x)
 \int_{0}^{\infty}  \mathrm{d} t \; \bigg[ {e^{-t \mu^2}
 \over 8 \pi t}
  - { e^{-t(H_0 -E)} \over 8 \pi t}
\bigg] \chi_{g}(x) + \mathrm{Regular} \; \mathrm{terms} \; \cr &\
& = {1 \over 8 \pi} \int_{\mathcal{M}} \mathrm{d}_g^2 x \;
\chi^{\dag}_{g}(x)\, \ln \left({H_0 -E \over \mu^2}\right) \,
\chi_{g}(x) + \mathrm{Regular} \; \mathrm{terms} \;,
\eeqs
\end{widetext}
where the other terms in the asymptotic expansion (\ref{asymheat})
do not give rise to an infinite result.

Let us give a better justification of this choice: we will again
assume that the orthofermion operators act on some smooth functions;
since the set of smooth functions are dense in the Hilbert space
norm, this is allowed. We will write one of the heat kernels as a
distributional solution in (\ref{K^2}), and use the fact that
$-\nabla^2_g$ is a self adjoint operator,
\begin{widetext}
\beqs
&\ & \int_{\mathcal{M}^2} \mathrm{d}_g^2 y \, \mathrm{d}_g^2
x \; \psi_a^*(x) \, K_{t}(x,y;g) e^{t \nabla^{2}_{g}} \,
\delta_g^{(2)}(x,y) e^{-t(H_0-E)} \psi_b(y) \cr &\ & =
\int_{\mathcal{M}^2} \mathrm{d}_g^2 y \, \mathrm{d}_g^2 x \;
\bigg[ e^{t \nabla^2_{g}} \psi_a^*(x) \, K_{t}(x,y;g) \bigg] \,
\delta_g^{(2)}(x,y) e^{-t(H_0-E)} \, \psi_b(y) \;.
\label{beforegilkey}
\eeqs
\end{widetext}
Let us expand the exponential $e^{t \nabla^{2}_{g}}$ into a formal
power series and define
\beqs
\hskip-0.5cm (\nabla_g)^k :=
\left\{%
\begin{array}{ll}
    (\nabla_g^2)^{k/2}, & \hbox{if $k=0,2,4,6,\ldots$ ;} \\
    \nabla_g (\nabla^2_g)^{(k-1)/2}, & \hbox{if $k=1,3,5,7,\ldots$,} \\
\end{array}
\right. \label{nabla def}
\eeqs
where $(\nabla_g f)^i = g^{ij} {\partial f \over \partial x^j}$
for any smooth function $f$ on $\mathcal{M}$. Then we get terms of
the following form
\be
t^k \left[ (\nabla_{g})^k \, \psi_a^*(x) \right] \, t^{n-k} \,
\left[ (\nabla_{g})^{n-k} \, K_{t}(x,y;g) \right] \;.
\ee
As $t \rightarrow 0^+$, the most singular terms in this expansion
will come from the terms with the highest number of derivatives of
the heat kernel, thanks to the following theorem (Lemma 1.7.7 in
\cite{gilkey}): If $D^{\alpha}_{x}$ is a differential operator
(acting on the functions of variable $x$) of order $\alpha$, then
the asymptotic expansion of the kernel of the operator
$D^{\alpha}_{x} e^{t \nabla^2_{g}} $ on the diagonal (in $D$
dimensions)
\be
D^{\alpha}_{x} K_{t}(x,y;g) |_{x=y} \sim \sum_{k=0}^{\infty}
t^{-(D+\alpha-k)/2} e_{k}(x, D^{\alpha}_{x}, \nabla^2_{g})
\label{theoremgilkey} \;,
\ee
where $e_k$ are smooth local invariants of the jets of the symbols
of the operators $D^{\alpha}_{x}$ and $\nabla^2_{g}$. Also $e_k$
are zero if $k+\alpha$ is odd. Thus, the most singular terms will
come from the highest powers of the Laplacian acting on the heat
kernel when we formally expand the exponential operator. This
means that the dominant contribution to equation
(\ref{beforegilkey}) is given by
\begin{widetext}
\be
\int_{\mathcal{M}^2} \mathrm{d}_g^2 y \, \mathrm{d}_g^2 x \;
\psi_a^*(x) \, \left[e^{t \nabla^2_{g}} \, K_{t}(x,y;g) \right] \,
\delta_g^{(2)}(x,y) \, e^{-t(H_0-E)}\psi_b(y)\;.
\ee
If we make use of the heat equation (\ref{heatequation}) in the
above, we may infer that
\beqs
e^{t \nabla^2_{g}} \, K_{t}(x,y;g) = \left[e^{t {\partial
\over
\partial t'}} \, K_{t'}(x,y;g)\right]\bigg|_{t'=t} \;.
\eeqs
\end{widetext}
Using the fact that $e^{t{\partial \over \partial t'}}$ generates
a time translation by an amount $t$, which is again true in the
sense of distributions:
\beqs
& & \lim_{t' \rightarrow t} e^{t'{\partial \over \partial
t}} K_{t}(x,x';g) = \lim_{t' \rightarrow t} K_{t+t'}(x,x';g) =
K_{2t}(x,x';g) \;,
\eeqs
we see that the most singular part of the integral as $t
\rightarrow 0^+$ turns out to be
\be
\int_{\mathcal{M}} \mathrm{d}_g^2 y \; \psi_a^*(y) \,
K_{2t}(y,y;g) \, e^{-t(H_0-E)} \, \psi_b(y) \;,
\ee
where we have taken the integral with respect to $x$. This
justifies our choice of the coupling constant (\ref{coupling
constant}).

\subsection{Two Dimensional Flat Case Revisited}

We can also explicitly show that this idea works for the same
model on flat space $\mathbb{R}^2$ by writing the principal
operator in momentum space that has already been calculated in
\cite{rajeevbound}. For this purpose, let us consider the first
part of equation (\ref{Phi}) in a two dimensional plane, i.e.,
\begin{widetext}
\beqs
&\ & \hskip-1cm  \int_{\mathbb{R}^4} \mathrm{d}^2 x \,
\mathrm{d}^2 x' \chi^{\dag}(\mathbf{x}) \lim_{\epsilon\rightarrow
0^+}
 \int_{\epsilon}^{\infty} \mathrm{d} t \; \bigg[ {e^{- t\mu^2}
 \over 8 \pi t}
\delta^{(2)}(\mathbf{x},\mathbf{x}') -
K_{t}^{2}(\mathbf{x},\mathbf{x}') e^{-t(H_0 -E)} \bigg]
\chi(\mathbf{x}') \;. \label{flatbeforeheat}
\eeqs
\end{widetext}
Substituting the explicit form of the heat kernel in
$\mathbb{R}^2$ \cite{chavel2}
\be
K_{t}(\mathbf{x},\mathbf{x}')= {e^{-|\mathbf{x}-\mathbf{x}'|^2
/4 t} \over 4 \pi t} \;,
\ee
we find for (\ref{flatbeforeheat})
\begin{widetext}
\beqs
&\ & \int_{\mathbb{R}^4} \mathrm{d}^2 x \, \mathrm{d}^2 x'
\; \chi^{\dag}(\mathbf{x}) \lim_{\epsilon\rightarrow 0^+}
 \int_{\epsilon}^{\infty} \mathrm{d} t \; \bigg[ {e^{- t \mu^2 }
 \over 8 \pi t}
\delta^{(2)}(\mathbf{x},\mathbf{x}') - \;
{e^{-|\mathbf{x}-\mathbf{x}'|^2 /4 t} \over (4 \pi t)} {e^{-
|\mathbf{x}-\mathbf{x}'|^2 /4 t} \over (4 \pi t)} e^{-t (H_0 -E)}
\bigg] \chi(\mathbf{x}') \;.
\eeqs
If we write the heat kernel as a Fourier transform of a function
$e^{-t \mathbf{p}^2}$ and then change the integration order above
for the second term, we get
\beqs
&\ & \hskip-1cm  \int_{\mathbb{R}^2} \mathrm{d}^2 x \,
\mathrm{d}^2 x' \chi^{\dag}(\mathbf{x})
 \int_{\epsilon}^{\infty} \mathrm{d} t \; {e^{- |\mathbf{x}-\mathbf{x}'|^2 /4 t}
\over (4 \pi t)} {e^{-|\mathbf{x}-\mathbf{x}'|^2 /4 t} \over (4
\pi t)} e^{-t(H_0 -E)} \chi(\mathbf{x}') \cr &\ & =
 \int_{\mathbb{R}^4} \mathrm{d}^2 x \, \mathrm{d}^2 x' \; \chi^{\dag}(\mathbf{x})
 \int_{\epsilon}^{\infty} \mathrm{d} t \; \int {\mathrm{d}^2 p \over (2 \pi)^2}
{e^{i \mathbf{p}.(\mathbf{x}-\mathbf{x}')-t \mathbf{p}^2 /2} \over
(8 \pi t)}  e^{-t(H_0 -E)} \chi(\mathbf{x}') \cr &\ & =
\int_{\mathbb{R}^2} {\mathrm{d}^2 p \over (2 \pi)^2}
\chi^{\dag}(\mathbf{p}) \int_{\epsilon}^{\infty} \mathrm{d} t \;
{e^{-t (H_0 -E + \mathbf{p}^2 /2)} \over (8 \pi t) }
\chi(\mathbf{p}) \;.
\eeqs
Then, equation (\ref{flatbeforeheat}) becomes as $\epsilon
\rightarrow 0^+$
\beqs
\hskip-0.4cm \int_{\mathbb{R}^2} {\mathrm{d}^2 p \over (2
\pi)^2} \chi^{\dag}(\mathbf{p}) \int_{\epsilon}^{\infty}
\mathrm{d} t \; \Bigg[ {e^{-t \mu^2}
 \over 8 \pi t} -  {e^{-t (H_0 -E + \mathbf{p}^{2}
 /2)} \over 8 \pi t} \Bigg] \chi(\mathbf{p}) = {1 \over 8 \pi} \int_{\mathbb{R}^2} {\mathrm{d}^2 p
\over (2 \pi)^2} \chi^{\dag}(\mathbf{p}) \ln \left({ H_0 - E +
\mathbf{p}^2 /2 \over \mu^2 }\right)\chi(\mathbf{p}) \;.
\eeqs
\end{widetext}
This is exactly the same result that was already calculated for
this model defined in the flat space $\mathbb{R}^2$
\cite{rajeevbound}.

\subsection{Analysis of the Bound State Spectrum} \label{Analysis of the Bound State Spectrum}

As a result of our analysis, we now have obtained a finite well defined model, that is,
the resolvent of the system is expressed in terms of the
well-defined principal operator given in (\ref{Phi})
\begin{widetext}
\beqs
R(E) = {1 \over H_0 -E} + {1 \over 2} {1 \over H_0 -E} \,
\int_{\mathcal{M}} \mathrm{d}_g^3 y \; \phi_{g}^{\dag}(y)
\phi_{g}^{\dag}(y) \chi_{g}(y) \, \Phi^{-1}(E) \,
\int_{\mathcal{M}} \mathrm{d}_g^3 y \; \phi_{g}(y) \phi_{g}(y)
\chi_{g}^{\dag}(y)\, {1 \over H_0 -E} \;.
\label{resolventlambdaphi4}
\eeqs
\end{widetext}
This is the analogue of the Krein's formula in the case of the
many-body version of the point interactions. All the information
of the spectrum of the problem can be determined from the above
resolvent operator. In this subsection we will discuss the spectral properties of our model, especially bound state
spectrum.

The poles in the resolvent corresponds to
the bound states. For non-compact manifolds, there can not be any pole due to the free
resolvent. For compact manifolds, we are interested in the poles below the poles of the free resolvent. These imply that the roots of the principal operator (\ref{Phi})
\be
\Phi(E)|\Psi \rangle = 0 \;, \label{spectrum}
\ee
determines the possible bound state spectrum. As in the case of the problem
where the particles only interact with an external Dirac delta
potential, which  displays a dimensional transmutation in two
dimensions \cite{huang, thorn, Camblong}, our model constructed
above also realizes a kind of dimensional transmutation. This can
be seen as follows. From the original Hamiltonian (\ref{ham}) that
we have started, it is easy to see that the coupling constant is
dimensionless so that there seems to be no parameter whatsoever to
yield an estimate of the energy by naive dimensional analysis.
However, if we have a  length scale coming from the geometry, such
as the curvature, this provides a geometric energy scale which is
there also for the free theory. Nevertheless, even if it is the
case, a new dimensional parameter $\mu^2$ shows up after the
renormalization procedure from the relation (\ref{coupling
constant}). Therefore, we can say that this is a general
dimensional transmutation and it is most striking when there is no
intrinsic energy scale coming from the geometry
\cite{ermanturgut}.

\begin{figure}[h!]
  \includegraphics[width=10cm]{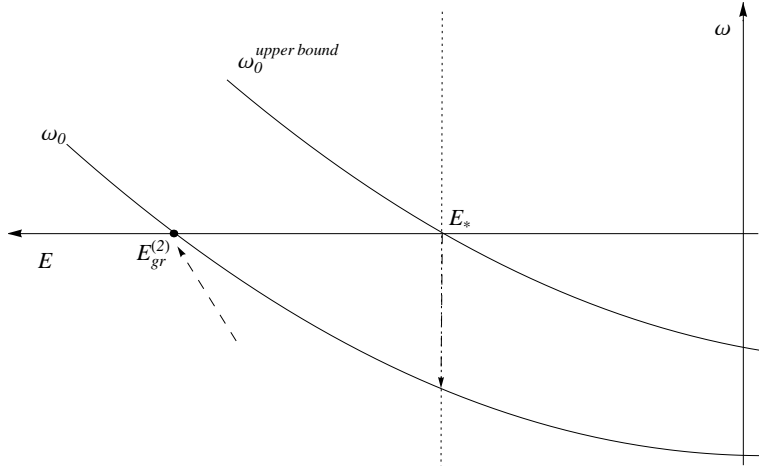}\\
  \caption{A Typical Flow of the First Eigenvalue of Principal Operator.}\label{omegaflow0}
\end{figure}

After the renormalization of the coupling constant, we must be
able to predict the other measurable quantities in terms of the
measured two-particle bound state energy  $E^{(2)}_{gr}$, in our version
the arbitrary scale $-\mu^2$ should be solved in favor of this binding energy. In flat space $\mathbb{R}^2$, two-body
solution is given by $E^{(2)}_{gr}=-\mu^2$ \cite{rajeevbound}. {\it From
this point on we assume $-\mu^2$ is expressed in terms of $E^{(2)}_{gr}$}.
We make the following comment, let us consider a
compact manifold, and apply the variational principle for \textit{the
first eigenvalue} $\omega_0(E)$ of $\Phi(E)$ in \textit{the two-boson
sector}. Since we are on a compact manifold we choose the orthofermion
wave function as constant, ${1\over \sqrt{V(\mathcal{M})}}$. We
now calculate the expectation value of the principal operator $\Phi(E)$ by the following variational ansatz
\be
| \Psi^{var} \rangle = |0 \rangle \otimes {1 \over \sqrt{V({\mathcal{M}}})}  \int_{\mathcal{M}} \mathrm{d}_{g}^{3} x \; \chi^{\dagger}(x) |0 \rangle \;.
\ee
Since $\Phi(E)$ is normal ordered, all the parts which contain bosonic creation and annihilation operators will vanish. The only term which survives
sets an upper bound for $\omega_0(E)$. Hence,
\begin{widetext}
\beqs
\omega_0(E) & \leq & \langle \Psi^{var} | \Phi(E) |\Psi^{var} \rangle \cr & \leq & \int_0^\infty \mathrm{d}t \; \bigg[{e^{-\mu^2
t}\over 8\pi t}-\bigg({1\over V(\mathcal{M})}\int_{\mathcal{M}^2}
\mathrm{d}_{g}^{2} x \;\mathrm{d}_{g}^{2} x' \;   K_{t}(x,x';g) K_{t}(x',x;g) \bigg)e^{-|E|t} \bigg] \cr & \leq & \int_0^\infty \mathrm{d}t \; \bigg[{e^{-\mu^2
t}\over 8\pi t}-\bigg({1\over V(\mathcal{M})}\int_{\mathcal{M}}
\mathrm{d}_{g}^{2} x \; K_{2t}(x,x;g)\bigg)e^{-|E|t} \bigg] \;, \label{omega0variational}
\eeqs
\end{widetext}
where we have used the semi-group property of the heat kernel (\ref{semigroupprop}). Compactness of the manifold implies that
it is complete as a Riemannian manifold and it has a Ricci tensor bounded from below which we formally write $Rc \geq \kappa$.
As a result of the theorem proven by J. Cheeger and S.-T. Yau \cite{CheegerYau}, the heat kernel has the following lower bound
\be
K_t(x,y;g) \geq K_{t}^{\kappa}(d_g(x,y)) \;, \label{lowerbound heatkernel}
\ee
where $K_{t}^{\kappa}$ is the heat kernel of the simply connected complete two dimensional manifold of constant
sectional curvature $\kappa$. In particular, we choose $K_{t}^{\kappa}(d_g(x,y))$ as the heat kernel of the two dimensional Hyperbolic manifold
$\mathbb{H}^2$ for $\kappa=-1/R^2$, where $R$ is the corresponding length scale. In case the lower bound is positive we may choose the heat kernel
for the two dimensional flat space and the argument below becomes even simpler. Since the heat kernel for two dimensional
Hyperbolic manifolds is explicitly known \cite{Grigoryan}, a lower bound of the diagonal heat kernel in (\ref{omega0variational}) is
\be
K_{2t}(x,x) \geq {R \sqrt{2} \over (8 \pi t)^{3/2}} \; e^{-t/2R^2} \int_{0}^{\infty} \mathrm{d}s \; {s \; e^{-s^2 R^2/8t} \over \sqrt{\cosh s-1}} \;.
\ee
From the expansion of the function $\cosh$, we can write the denominator as $\sqrt{\cosh s-1}= \sqrt{\sum_{k=1}^{\infty} s^{2k}/(2k)!} =
({s/\sqrt{2}})\sqrt{\sum_{k=1}^{\infty} 2 s^{2k-2}/(2k)!}$. Then we have
\beqs
\omega_0(E) & \leq & \int_0^\infty \mathrm{d}t \; \bigg[{e^{-\mu^2
t}\over 8\pi t} + {2 R \over (8 \pi t)^{3/2}} \; e^{-t/2R^2} \int_{0}^{\infty} \mathrm{d}s \; e^{-s^2 R^2/8t-|E|t}
\left(-1+ \left( 1- {1 \over \sqrt{\sum_{k=1}^{\infty} 2 s^{2k-2}/(2k)!}} \right)\right)\bigg] \;,
\eeqs
where we have added and subtracted $1$ in the parenthesis above. Since $\sum_{k=1}^{\infty} 2 s^{2k-2}/(2k)! \leq e^{s^2}$ for all $s\geq 0$, we have
\beqs
\omega_0(E) & \leq & {1 \over 8 \pi} \ln \left({|E|+{R^2 \over 2} \over \mu^2}\right) + \int_{0}^{\infty} \mathrm{d} t \;
{1 \over 4 \pi t} \left( 1- {1 \over \sqrt{1+4t}}\right) e^{-t(|E|+{R^2 \over 2})} \;.
\eeqs
Using $\sqrt{1+4t} \leq 1 +2t$ and $1+2t \geq 1$ for all $t\geq 0$, we get
\beqs
\omega_0(E) & \leq & {1 \over 8 \pi} \ln \left({|E|+{R^2 \over 2} \over \mu^2}\right) + {1 \over 2\pi} {1 \over |E|+{R^2 \over 2}}\;.
\eeqs
For large values of $\mu^2$ there always exists a unique $E_* <0$ such that
\be
{1 \over 8 \pi} \ln \left({|E_*|+{R^2 \over 2} \over \mu^2}\right) = -{1 \over 2\pi} {1 \over |E_*|+{R^2 \over 2}} \label{twobodyEgrmu}\;.
\ee
As we will prove in this section
\be
{\partial \omega_0 \over \partial E} <0 \;,
\ee
to get the true
zero $E^{(2)}_{gr}$ of $\omega_0(E)$, we must further decrease $E$ (or increase $|E|$)
so that we will have a well-defined expression of $\mu^2$ in terms of
two-particle binding energy $E^{(2)}_{gr}< E_*<0$, as shown in Figure \ref{omegaflow0}.
Therefore, by assuming that two-body problem is solved, we can
then study n-body problem.  Since we are only interested in the
bound states of the model at the moment, we should be able to
determine $n$-particle bound states after the renormalization
procedure. The exact treatment of this problem is rather
difficult.  Assuming that the details of the two-body interaction can
be understood,  we will study the model in the mean field
approximation in Section \ref{Mean Field App}.
Before embarking on studying the mean field analysis, we will make
some general remarks about the bound state spectrum of the
problem.

It is a well known fact that the residue of the resolvent at its
isolated pole $\mu$ is the projection operator $\mathbb{P}_\mu$ to
the corresponding eigenspace of the Hamiltonian
\be
\label{projection resolvent}
\mathbb{P}_{\mu}= -{1 \over 2 \pi i} \oint_{\Gamma_{\mu}}
\mathrm{d} E \; R(E)\;,
\ee
where $\Gamma_{\mu}$ is a small contour enclosing the isolated
eigenvalue $\mu$ in the complex energy plane \cite{simon}. Let us
suppose that there exist a ground state and choose our contour
enclosing this ground state energy, namely $E_{gr}$. Then, the
above integral of $R(E)$ gives the projection to the eigenspace
$|\Psi_0 \rangle \langle \Psi_0|$ corresponding to the minimum
eigenvalue of the renormalized Hamiltonian for the many-body system.

From (\ref{Phi}), it is easily seen that the principal operator
formally satisfies $\Phi^{\dag}(E)= \Phi(E^{*})$. We assume that
$\Phi(E)$ defines  a self-adjoint holomorphic family of type A
\cite{kato}, so that we can apply the spectral theorem for the
principal operator or inverse of it. Since the principal operator
$\Phi(E)$ acts on $\mathcal{F_B}^{(n-2)} \otimes
\mathcal{L}^2(\mathcal{M})$, we have
\beqs
& & \Phi^{-1}(E) = \sum_{k} {1 \over \omega_k(E)}
\mathbb{P}_k(E) + \int_{\sigma} \mathrm{d} \omega(E) \; {1 \over
\omega(E)} \mathbb{P}_{\omega}(E) \;, \label{Phiinverse}
\eeqs
where the projection operator
\begin{widetext}
\beqs
\mathbb{P}_k(E)  = |\phi_k(E)\rangle \langle \phi_k(E)| =
|\omega_k(E); \Omega_k(E) \rangle \langle \omega_k(E);
\Omega_k(E)| \;,
\eeqs
\end{widetext}
is given in terms of $n-2$ bosonic particle state and one-particle
orthofermion state:
\begin{widetext}
\beqs
|\omega_k \rangle & = & \int_{\mathcal{M}^{n-2}} \mathrm{d}^{2}_{g}x_1 \, \mathrm{d}^{2}_{g}x_2 \ldots \mathrm{d}^{2}_{g}x_{n-2} \;
u_k(x_1,\ldots,x_{n-2}) |x_1 \ldots x_{n-2} \rangle \cr
|\Omega_k \rangle & =& \int_{\mathcal{M}} \mathrm{d}^{2}_{g}x \; \psi_k(x) \chi^{\dag}_{g}(x)|0 \rangle \;.
\eeqs
\end{widetext}
Here $\omega_k(E)$ and $|\omega_k(E);
\Omega_k(E)\rangle$ are the eigenvalues and
the eigenvectors of the principal operator, respectively. Similarly, the (generalized) projection operator
\begin{widetext}
\beqs
\mathbb{P}_{\omega}(E) &=&  |\phi(E) \rangle
\langle \phi(E)| = |\omega(E); \Omega(E) \rangle \langle
\omega(E); \Omega(E)|
\eeqs
\end{widetext}
corresponds to the continuous eigenvalues and eigenvectors of the principal operator. We
assume that the principal operator has discrete as well as
continuous eigenvalues and the bottom of the spectrum corresponds
to a non-degenerate eigenvalue. Above integral is taken over the
continuous spectrum $\sigma(\Phi)$ of the principal operator (for
simplicity, we write it formally, it should be written more
precisely as a Riemann-Stieltjies integral).

\begin{figure}[h!]
  \includegraphics[width=10cm]{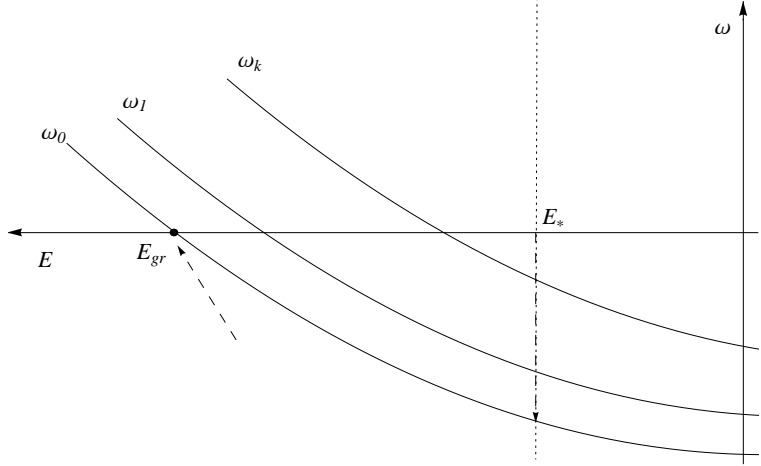}\\
  \caption{A Typical Flow of the Eigenvalues of Principal Operator.}\label{omegaflow1}
\end{figure}

As emphasized in the previous section, the bound state spectrum
corresponds to the solutions of the zero eigenvalues of the
principal operator (\ref{Phi}). In order to estimate the ground
state energy of our system, it is crucial to determine how the
eigenvalues $\omega_k$ evolve with $E$. For this purpose, let us
calculate the derivative of the eigenvalue $\omega_k$ of the
principal operator with respect to $E$. If we apply the
Feynman-Hellman theorem to the eigenvalue problem for the
principal operator, we get
\beqs
{\partial \omega_k \over \partial E} = \langle \phi_k |
{\partial \Phi(E) \over \partial E} | \phi_k \rangle = \bigg\langle
{\partial \Phi(E) \over \partial E} \bigg\rangle  \;.
\label{feynmanhellman}
\eeqs
A direct computation for the derivative of the principal operator
(\ref{Phi}) with respect to the energy $E$ gives
\begin{widetext}
\beqs
&\ & {\partial \Phi(E) \over \partial E} = -
\Bigg[\int_{\mathcal{M}^2} \mathrm{d}_g^2 x \, \mathrm{d}_g^2 x'
\chi^{\dag}_{g}(x) \int_{0}^{\infty} \mathrm{d} t\; t \;
K_{t}^{2}(x,x';g) e^{-t(H_0 -E)} \chi_{g}(x') + {1 \over 2}
\int_{\mathcal{M}^6} \mathrm{d}_g^2 x \,\mathrm{d}_g^2 x' \,
\mathrm{d}_g^2 x_1 \, \mathrm{d}_g^2 x_2 \, \mathrm{d}_g^2 x'_1 \,
\mathrm{d}_g^2 x'_2 \cr & & \times \;
\chi^{\dag}_{g}(x)\,\chi_{g}(x') \, \phi^{\dag}_{g}(x'_1)\,
 \phi^{\dag}_{g}(x'_2)
\int_{0}^{\infty} \mathrm{d} t\; t \; K_{t}(x_1,x;g) \,
 K_{t}(x_2, x;g) \, K_{t}(x',x'_1;g) \,
 K_{t}(x',x'_2;g) \, e^{-t (H_0 -E)}
\phi_{g}(x_1)\, \phi_{g}(x_2)\cr &\ & \hspace{3cm} + 2
\int_{\mathcal{M}^4} \mathrm{d}_g^2 x \,\mathrm{d}_g^2 x' \,
\mathrm{d}_g^2 x_1 \, \mathrm{d}_g^2 x_2 \;
\chi^{\dag}_{g}(x)\,\chi_{g}(x') \, \phi^{\dag}_{g}(x_1) \ \cr &\
& \hspace{6cm} \times \int_{0}^{\infty} \mathrm{d} t \;t \,
K_{t}(x_2,x;g) \,
 K_{t}(x,x';g)\, K_{t}(x',x_1;g)\, e^{-t (H_0 -E)} \phi_{g}(x_2)
\Bigg] \;. \label{derivativeofPhi}
\eeqs
For simplicity, we will separate the terms in the expectation
value of the principal operator in (\ref{feynmanhellman}), using
(\ref{derivativeofPhi}). Let us first consider the first term
\beqs
\int_{\mathcal{M}^2} \mathrm{d}_g^2 x \, \mathrm{d}_g^2 x'
\; \psi^{*}(x) \int_{0}^{\infty} \mathrm{d} t\; t \;
K_{t}^{2}(x,x';g) \langle \omega_k |e^{-t(H_0 -E)}|\omega_k
\rangle \psi(x') \;,
\eeqs
\end{widetext}
where $\psi(x)$ is the wave function of the \textit{orthofermion}. If we
think of the factor $t$ in the above integrand as an integral
$\int_{-t}^{t} (\mathrm{d} u/2)$ and then make the change of
variables $t=t_1+t_2$, $u=t_1-t_2$, we readily obtain
\begin{widetext}
\beqs
\int_{\mathcal{M}^2} \mathrm{d}_g^2 x \, \mathrm{d}_g^2 x'
\; \psi^{*}(x) \int_{0}^{\infty} \int_{0}^{\infty} \; \mathrm{d}
t_1 \, \mathrm{d} t_2 \;  \; K_{t_1+t_2}^{2}(x,x';g) \langle
\omega_k |e^{-(t_1+t_2)(H_0 -E)}|\omega_k \rangle \psi(x') \;.
\label{1stderivative1}
\eeqs
Using the semi-group property of the heat kernel
(\ref{semigroupprop}), equation (\ref{1stderivative1}) can be
rewritten as
\beqs
& & \hskip-1cm \int_{\mathcal{M}^2} \mathrm{d}_g^2 x \,
\mathrm{d}_g^2 x' \; \psi^{*}(x) \int_{0}^{\infty}
\int_{0}^{\infty} \; \mathrm{d} t_1 \, \mathrm{d} t_2 \;
\int_{\mathcal{M}^2} \mathrm{d}_g^2 z_1 \, \mathrm{d}_g^2 z_2 \;
K_{t_1}(x,z_1;g) K_{t_2}(z_1;x';g) K_{t_1}(x,z_2;g) \cr  & &
\hspace{5cm} \times \; K_{t_2}(z_2,x';g) \langle \omega_k
|e^{-(t_1+t_2)(H_0 -E)}|\omega_k \rangle \psi(x') \;.
\eeqs
Changing the order of integrations we find
\beqs
\int_{\mathcal{M}^2} \mathrm{d}_g^2 z_1 \, \mathrm{d}_g^2
z_2 \; \Bigg|\Bigg| \int_{0}^{\infty} \; \mathrm{d} t_1 \;
e^{-t_1(H_0 -E)} \int_{\mathcal{M}} \mathrm{d}_g^2 x \;
K_{t_1}(z_1,x;g) K_{t_1}(x,z_2;g) \psi^{*}(x) | \omega_k \rangle
\Bigg|\Bigg|^2 \;, \label{1stderivative}
\eeqs
\end{widetext}
which is obviously always positive. Now we return to the
expectation value of the second and third terms in
(\ref{derivativeofPhi}). It is easy to see that they can be
expressed as
\begin{widetext}
\beqs
& & {1 \over 2} \int_{0}^{\infty} \mathrm{d} t \; t \,
\Bigg|\Bigg| \int_{\mathcal{M}^3} \mathrm{d}_g^2 x_1 \,
\mathrm{d}_g^2 x_2 \, \mathrm{d}_g^2 x \; K_{t}(x_1,x;g)
K_{t}(x,x_2;g) \psi^{*}(x) \; e^{-{t \over 2}(H_0-E)} \phi_g(x_1)
\phi_g(x_2)|\omega_k \rangle \Bigg|\Bigg|^2 \cr & & \hskip-0.5cm +
\; 2 \int_{\mathcal{M}} \mathrm{d}_g^2 z \int_{0}^{\infty}
\mathrm{d} t \; t \, \Bigg|\Bigg| \int_{\mathcal{M}^2}
\mathrm{d}_g^2 x \, \mathrm{d}_g^2 x_2 \; K_{t/2}(z,x;g)
K_{t}(x,x_2;g) \psi^{*}(x) \; e^{-{t \over 2}(H_0-E)}
\phi_g(x_2)|\omega_k \rangle \Bigg|\Bigg|^2
\;,\label{2nd3thderivative}
\eeqs
\end{widetext}
where we have used the fact that we can rewrite the second heat
kernel $K_t(x,x';g)$ in the third term of (\ref{derivativeofPhi})
as $\int_{\mathcal{M}} \mathrm{d}_{g}^{2} z \; K_{t/2}(x,z;g)
K_{t/2}(z,x';g)$ by the semi-group property (\ref{semigroupprop}).
Consequently, we obtain
\beqs
{\partial \omega_k \over \partial E} < 0 \;.
\label{derivative of eigenvalue phi}
\eeqs
The eigenvalues $\omega_k(E)$'s flow with $E$ in accordance with
(\ref{derivative of eigenvalue phi}), that is, these are
monotonically decreasing functions of $E$. For sufficiently small
values of $E$, there can not be a zero eigenvalue of the principal
operator since the energy must be bounded from below. Moreover,
for a given $E_{*}$ the eigenvalues can be ordered as
$\omega_0(E_*)<\omega_1(E_*)<\ldots $. Therefore, due to
(\ref{derivative of eigenvalue phi}) and non-degeneracy of the
lowest eigenvalue $\omega_0$, only the minimum eigenvalue
$\omega_0$ flows to its zero value at the minimum energy
$E=E_{gr}$. Hence, the ground state corresponds to the zero of the
minimum eigenvalue $\omega_0(E)$ of $\Phi(E)$, as shown in Figure \ref{omegaflow1}.

\begin{figure}[h!]
  \includegraphics[width=10cm]{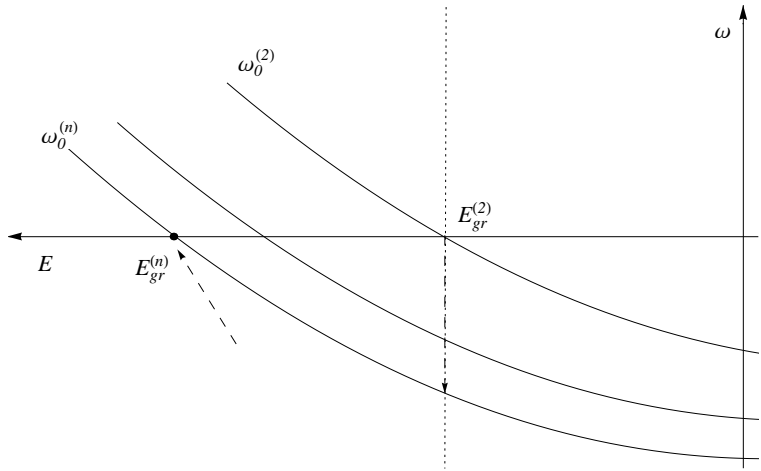}\\
  \caption{A Typical Flow of the Eigenvalues of Principal Operator in the two and $n$-boson sector.} \label{omegaflow2}
\end{figure}

We may now show that $E^{(n)}_{gr} \leq E^{(2)}_{gr}$ for compact manifolds. To see this, we take the solution of two-body ground state $|\Psi^{(2)} \rangle $ as
\beqs
|\Psi^{(2)} \rangle = |0 \rangle
\otimes \int_{\mathcal{M}} \mathrm{d}_{g}x \; \psi_{0}^{(2)}(x) \chi^{\dag}(x)|0\rangle
\eeqs
and then make a new ansatz $|\Psi^{var} \rangle$ for the $n$-body problem in the form
\be
|\Psi^{var} \rangle = {1 \over [V(\mathcal{M})]^{(n-2)/2}} \int_{\mathcal{M}^{n-2}} \mathrm{d}_{g}x_1 \;\ldots \mathrm{d}_{g}x_{n-2} |x_1, \ldots, x_{n-2} \rangle
\otimes \int_{\mathcal{M}} \mathrm{d}_{g}x \; \psi_{0}^{(2)}(x) \chi^{\dag}(x)|0\rangle \;.
\ee
It is convenient to split the principal operator $\Phi(E)$ given in (\ref{Phi}) as $K(E)-U(E)$, where
\be
K(E)=
\int_{\mathcal{M}^2} \mathrm{d}_g^2 x \,
\mathrm{d}_g^2 x' \chi^{\dag}_{g}(x)
 \int_{0}^{\infty} \mathrm{d} t \; \bigg[ {e^{- t \mu^2 }
 \over 8 \pi t}
\delta_g^{(2)}(x,x')  - K_{t}^{2}(x,x';g) e^{-t(H_0 -E)} \bigg]
\chi_{g}(x')
\ee
and
\begin{widetext}
\beqs
& & \hskip-2cm U(E)={1 \over 2} \int_{0}^{\infty} \mathrm{d} t \; \bigg[ \int_{\mathcal{M}^3} \mathrm{d}_g^2 x \, \mathrm{d}_g^2 x_1 \, \mathrm{d}_g^2 x_2 \,
K_{t}(x_1,x;g) \, K_{t}(x,x_2;g) \,\phi_{g}(x_1)\, \phi_{g}(x_2) \, \chi_{g}(x)\bigg]^{\dag}  e^{-t (H_0 -E)} \cr & & \hspace{3cm}
\times \; \bigg[ \int_{\mathcal{M}^3} \mathrm{d}_g^2 x'\, \mathrm{d}_g^2 x'_1 \, \mathrm{d}_g^2 x'_2 \,
K_{t}(x'_1,x';g) \, K_{t}(x',x'_2;g) \,\phi_{g}(x'_1)\, \phi_{g}(x'_2) \, \chi_{g}(x')\bigg] \cr & & + \, 2 \int_{\mathcal{M}} \mathrm{d}_g^2 z \; \int_{0}^{\infty} \mathrm{d}t \;
\bigg[\int_{\mathcal{M}^2} \mathrm{d}_g^2 x \, \mathrm{d}_g^2 x_1 \,K_{t}(x_1,x;g) \, K_{t/2}(x,z;g) \,\phi_{g}(x_1) \chi_{g}(x)\bigg]^{\dag} e^{-t (H_0 -E)} \cr & & \hspace{3cm}
\times \; \bigg[\int_{\mathcal{M}^2} \mathrm{d}_g^2 x' \, \mathrm{d}_g^2 x'_1 \,K_{t}(x'_1,x';g) \, K_{t/2}(x',z;g) \,\phi_{g}(x'_1) \chi_{g}(x')\bigg] \label{U}\;,
\eeqs
\end{widetext}
where we have used the semi-group property of heat kernel (\ref{semigroupprop}) and the assumption that we can interchange the order of integrations.
By the variational principle,
\beqs
\omega_{0}^{(n)} (E_{gr}^{(2)}) \leq \langle \Psi^{var} | \Phi(E_{gr}^{(2)}) | \Psi^{var} \rangle =- \langle \Psi^{var}
| U(E_{gr}^{(2)}) |\Psi^{var} \rangle \;,
\eeqs
where $\langle \psi_{0}^{(2)} |K(E^{(2)}_{gr}) |\psi_{0}^{(2)} \rangle =0$. In order to calculate the above expectation value, we first show that
\beqs
& &
e^{-{t \over 2} (H_0 -E)} \int_{\mathcal{M}^3} \mathrm{d}_g^2 x'\, \mathrm{d}_g^2 x'_1 \, \mathrm{d}_g^2 x'_2 \,
K_{t}(x'_1,x';g) \, K_{t}(x',x'_2;g) \,\phi_{g}(x'_1)\, \phi_{g}(x'_2) \, \chi_{g}(x') |\Psi^{var} \rangle \cr & & =
{(n-2)^{1/2}(n-3)^{1/2} \over [V(\mathcal{M})]^{(n-2)/2}} e^{-{t \over 2}|E_{gr}^{(2)}|} \int_{\mathcal{M}^3} \mathrm{d}_g^2 x'\, \mathrm{d}_g^2 x'_1 \, \mathrm{d}_g^2 x'_2 \,
K_{t}(x'_1,x';g) \, K_{t}(x',x'_2;g) \psi_{0}^{(2)}(x') \cr & & \hspace{5cm} \times \; \int_{\mathcal{M}^{n-4}} \mathrm{d}_{g}y_3 \;\ldots \mathrm{d}_{g}y_{n-2} |y_{3}, \ldots, y_{n-2} \rangle
\cr & & = {(n-2)^{1/2}(n-3)^{1/2} \over [V(\mathcal{M})]^{(n-2)/2}} e^{-{t \over 2}|E_{gr}^{(2)}|} \int_{\mathcal{M}} \mathrm{d}_g^2 x' \,
\psi_{0}^{(2)}(x') \int_{\mathcal{M}^{n-4}} \mathrm{d}_{g}y_3 \;\ldots \mathrm{d}_{g}y_{n-2} |y_{3}, \ldots, y_{n-2} \rangle \;,
\eeqs
where we have used the fact that the free Hamiltonian operates on bosonic states and gives zero for constant wave functions.
We have also used the stochastic completeness of the heat kernel in the last line.
Then, the expectation value of the first term in (\ref{U}) becomes
\beqs
& &
{(n-2)(n-3) \over 2} \int_{0}^{\infty} \mathrm{d} t \; e^{-t|E_{gr}^{(2)}|} \bigg| \int_{\mathcal{M}} \mathrm{d}_g^2 x \,
\psi_{0}^{(2)}(x)\bigg|^2 {[V(\mathcal{M})]^{(n-4)} \over [V(\mathcal{M})]^{(n-2)}} \;,
\eeqs
which is finite due to
\beqs
\bigg|\int_{\mathcal{M}} \mathrm{d}_g^2 x \,
\psi_{0}^{(2)}(x)\bigg|^2 \leq
\bigg[\int_{\mathcal{M}} \mathrm{d}_g^2 x \,
|\psi_{0}^{(2)}(x)|^2 \bigg] \bigg[\int_{\mathcal{M}} \mathrm{d}_g^2 x \bigg] = V(\mathcal{M}) \;.
\eeqs
Similarly, we can calculate the expectation value of the second term in (\ref{U}) and get
\beqs
{2(n-2) \over V(\mathcal{M})} \int_{0}^{\infty} \mathrm{d} t \; \Bigg[\int_{\mathcal{M}} \mathrm{d}_g^2 z \,
\bigg|\int_{\mathcal{M}} \mathrm{d}_g^2 x \, K_{t/2}(x,z;g) \psi_{0}^{(2)}(x) \bigg|^2
\Bigg] e^{-t|E_{gr}^{(2)}|} \label{Kpsi} \;.
\eeqs
It can be shown that the above integral is finite if we use the eigenfunction expansions (\ref{heat kernel}) and (\ref{heatkernelexpansion}), so that we have
\beqs
\int_{\mathcal{M}} \mathrm{d}_g^2 z \,
\bigg|\int_{\mathcal{M}} \mathrm{d}_g^2 x \, K_{t/2}(x,z;g) \psi_{0}^{(2)}(x) \bigg|^2 =
\Bigg[\int_{\mathcal{M}} \mathrm{d}_g^2 z \, \bigg|\bigg|e^{{t \over 2} \nabla_{g}^{2}}|\psi_{0}^{(2)}\rangle \bigg|\bigg|^2
\Bigg] = \sum_{l} e^{-t \sigma_l} |\tilde{\psi}_{0}^{(2)}(l)|^2 \;,
\eeqs
where $\psi_{0}^{(2)}(x)= \sum_l f_l(x;g) \tilde{\psi}_{0}^{(2)}(l)$. Since $ \sum_{l} e^{-t \sigma_l} |\tilde{\psi}_{0}^{(2)}(l)|^2 \leq  e^{-t \sigma_0} \sum_{l}  |\tilde{\psi}_{0}^{(2)}(l)|^2$
and the minimum eigenvalue $\sigma_0=0$ for compact manifolds, the above integral is bounded from above by $1$ and
so that equation (\ref{Kpsi}) is finite.
Hence, we obtain
\beqs
& &
\omega_{0}^{(n)} (E_{gr}^{(2)}) \leq - {(n-2)(n-3) \over 2 |E_{gr}^{(2)}| [V(\mathcal{M})]^2} \bigg| \int_{\mathcal{M}} \mathrm{d}_g^2 x \,
\psi_{0}^{(2)}(x)\bigg|^2  \cr & & \hspace{4cm} - {2(n-2) \over V(\mathcal{M})} \int_{0}^{\infty} \mathrm{d} t \; \Bigg[\int_{\mathcal{M}} \mathrm{d}_g^2 z \,
\bigg|\int_{\mathcal{M}} \mathrm{d}_g^2 x \, K_{t/2}(x,z;g) \psi_{0}^{(2)}(x) \bigg|^2
\Bigg] e^{-t|E_{gr}^{(2)}|} < 0 \;.
\label{negative omega}
\eeqs
As a consequence of (\ref{derivative of eigenvalue phi}) and (\ref{negative omega}), to find the zero of $\omega_0(E)$ in the $n$-particle sector
we must reduce $E$ below $E_{gr}^{(2)}$, as shown in Figure \ref{omegaflow2}. This completes the proof.
We will now calculate $n$-particle ground state wave function in terms of the solution
$|\phi_0 \rangle$ of $\Phi(E_{gr})|\phi_0 \rangle=0$.

Let us expand the
minimum eigenvalue $\omega_0(E)$ near the bound state energy
$E_{gr}$
\begin{widetext}
\be
\omega_0(E) = \omega_0(E_{gr}) + (E- E_{gr}) {\partial \omega_0(E)
\over
\partial E}\bigg|_{E_{gr}}+ \cdots  = (E- E_{gr}) {\partial \omega_0(E)
\over
\partial E}\bigg|_{E_{gr}}+ \cdots  \;.
\ee
Using this result and (\ref{Phiinverse}), equation
(\ref{projection resolvent}) yields
\beqs
& & {1 \over 2} (H_{0}-E_{gr})^{-1} \int_{\mathcal{M}}
\mathrm{d}_{g}^{2} x \; \phi^{\dag}_{g}(x) \phi^{\dag}_{g}(x)
\psi_0(x) \left(- {\partial \omega_0(E)\over
\partial E}\bigg|_{E_{gr}}  \right)^{-1}
|\omega_0(E_{gr}) \rangle \langle \omega_0(E_{gr}) | \cr & &
\hspace{9cm} \times \int_{\mathcal{M}} \mathrm{d}_{g}^{2} y \;
\phi_{g}(y) \phi_{g}(y) \psi^{*}_{0}(y) (H_{0}-E_{gr})^{-1} \;.
\label{residue resolvent}
\eeqs
\end{widetext}
We assume that there is no other  pole coming from $(H_0 -
E)^{-1}$ near $E_{gr}$, and no other terms for $k \neq 0$
contribute to the integral around $E=E_{gr}$. Let the eigenvector
of the principal operator corresponding to the ground state be
\begin{widetext}
\beqs
& & | \phi_0(E_{gr}) \rangle = \int_{\mathcal{M}^{n-2}}
\mathrm{d}_{g}^{2}x_1 \cdots \mathrm{d}_{g}^{2}x_{n-2} \;
u_{0}(x_1,\cdots, x_{n-2}) |x_1 \cdots x_{n-2} \rangle
\int_{\mathcal{M}} \mathrm{d}_{g}^{2}x \; \psi_0(x)
\chi^{\dagger}_{g}(x)|0\rangle \;. \label{ground state eigenvector
of Phi}
\eeqs
\end{widetext}
By using the eigenfunction expansion of the creation and the
annihilation operators and their commutation relations, we will
shift all creation operators $\phi^{\dag}_{g}(x)$ in (\ref{residue
resolvent}) coming from (\ref{ground state eigenvector of Phi}) to
the leftmost
\begin{widetext}
\beqs
& & {1\over H_0-E} \phi^{\dag}_{g}(x) \phi^{\dag}_{g}(x)
\phi^{\dag}_{g}(x_1)\cdots\phi^{\dag}_{g}(x_{n-2}) =
\int_{\mathcal{M}^{n}} \mathrm{d}_{g}^{2} y_1 \cdots
\mathrm{d}_{g}^{2} y_{n} \; \phi^{\dag}_{g}(y_1)\cdots
\phi^{\dag}_{g}(y_{n}) \cr & & \hspace{4cm} \times \int_0^\infty
\mathrm{d} t \; e^{-t(H_0-E)} \, K_{t}(y_1,x;g) K_{t}(y_2,x;g)
K_{t}(y_3,x_1;g) \cdots K_{t}(y_{n},x_{n-2};g) \;,
\label{normalorderingcreation}
\eeqs
and all annihilation operators $\phi_{g}(x)$ in (\ref{residue
resolvent}) coming from (\ref{ground state eigenvector of Phi}) to
the rightmost
\beqs
& & \phi_{g}(x)\phi_{g}(x) \phi_{g}(x_1) \cdots
\phi_{g}(x_{n-2}) {1\over H_0-E} = \int_{\mathcal{M}^{n}}
\mathrm{d}_{g}^{2} y_1 \cdots \mathrm{d}_{g}^{2} y_{n} \;
\int_0^\infty \mathrm{d} t \; e^{-t(H_0-E)}  \cr & & \hspace{4cm}
\times \; K_{t}(y_1,x;g) K_{t}(y_2,x;g) K_{t}(y_3,x_1;g) \cdots
K_{t}(y_{n},x_{n-2};g) \phi_{g}(y_1)\cdots \phi_{g}(y_{n}) \;,
\label{normalorderingannihilation}
\eeqs
\end{widetext}
which are the generalized versions of equations we first used in
\cite{nrleemodelonmanifold}. Therefore, from equation
(\ref{residue resolvent}), we read the state vector $|\Psi_0
\rangle$ of our many-body system in terms of the eigenstate $|\phi_0 \rangle$ of the principal operator
\begin{widetext}
\beqs
& & |\Psi_0 \rangle = \int_{\mathcal{M}^{n}}
\mathrm{d}_{g}^{2} y_1 \cdots \mathrm{d}_{g}^{2} y_{n} \;
\Psi_{0}(y_1, \ldots, y_{n}) |y_1\cdots y_{n} \rangle \cr & & = {1
\over \sqrt{2}} \int_{\mathcal{M}^{n}} \mathrm{d}_{g}^{2} y_1
\cdots \mathrm{d}_{g}^{2} y_n  \; \int_{\mathcal{M}^{n-1}}
\mathrm{d}_{g}^{2} x_1 \cdots \mathrm{d}_{g}^{2}
x_{n-2}\mathrm{d}_{g}^{2} x \; {1 \over n!} \sum_{\sigma \in [1
\cdots n]} \int_0^\infty \mathrm{d} t \; e^{-t |E_{gr}|}
K_{t}(y_{\sigma(1)},x;g) K_{t}(y_{\sigma(2)},x;g)
  \cr & & \hspace{1cm} \times \;
K_{t}(y_{\sigma(3)},x_1;g)\cdots K_{t}(y_{\sigma(n)},x_{n-2};g)\;
u_{0}(x_1,\cdots,x_{n-2}) \psi_{0}(x) \left(- {\partial
\omega_0(E)\over
\partial E}\bigg|_{E_{gr}}  \right)^{-1/2}
|y_{1}\cdots y_{n} \rangle \;, \label{psi0}
\eeqs
\end{widetext}
where the sum runs over all permutations $\sigma$ of $[123\ldots
n]$. Comparing equation (\ref{ground state eigenvector of Phi}) and equation (\ref{psi0}), we see that the state
$|\Psi_0 \rangle$ is a complicated convoluted integral of the eigenstate $|\phi_0 \rangle$ with the heat kernels.

\section{Mean Field Approximation}
\label{Mean Field App}

In standard quantum field theory, one expects that all the bosons
have the same wave function $u(x)$ for the limit of large number
of bosons, i.e., as $n\rightarrow \infty$ and the wave function of
the system has the product form of the one-particle wave
functions. However, due to the singular structure of our problem,
the wave function in (\ref{psi0}) can not have a product form in
the large $n$ limit. In order to see this, we scale
$t=t'/|E_{gr}|$. With a hindsight coming from the proof that the
lower bound of the ground state energy grows exponentially with
the number of bosons in flat space \cite{rajeevbound,rajeevdimock}
we may assume that $E_{gr}$ grows fast enough as $n$ increases. In
this case, all integrals of the heat kernels are peaked around
$y_{\sigma(k)}$. (This is clear from (\ref{heatkernel delta}) and
also from the stochastic completeness assumption). Then, all
integrals of $x_{l}$ are
\begin{widetext}
\be
\int_{\mathcal{M}}\mathrm{d}_{g}^{2} x_l \; K_{t/|E_{gr}|}(x_l
,y_{\sigma(l+1)}) u_0(x_1,\ldots,x_l,\ldots, x_{n-2}) \approx
u_0(x_1,\ldots,y_{\sigma(l+1)},\ldots, x_{n-2}) \;,
\ee
\end{widetext}
for $l=1,\ldots,n-2$ as $n\rightarrow \infty$ and similarly for
$x$ integral. Then, the state $|\Psi_0 \rangle$ becomes
\begin{widetext}
\beqs
& & |\Psi_0 \rangle \approx {1 \over \sqrt{2}}
\int_{\mathcal{M}^{n}} \mathrm{d}_{g}^{2} y_1 \cdots
\mathrm{d}_{g}^{2} y_n \; {1 \over n!} \sum_{\sigma \in [1 \cdots
n]} \int_0^\infty \mathrm{d} t \; e^{-t|E_{gr}|}
K_{t}(y_{\sigma(1)},y_{\sigma(2)};g) \cr & & \hspace{5cm} \times
\; u_{0}(y_{\sigma(3)},\cdots,y_{\sigma(n)})
\psi_0(y_{\sigma(2)})\left(- {\partial \omega_0(E)\over
\partial E}|_{E_{gr}}  \right)^{-1/2}
|y_{1}\cdots y_{n} \rangle \;. \label{psi1}
\eeqs
One can understand the singular nature of the wave function in this limiting form more easily. We pick any two bosons,
and transform them through our formalism into an orthofermion, with its wave function $\psi_0$ to be determined consistently.
This orthofermion wave function corresponding to the pairing, could be quite regular, yet its multiplication  with the heat kernel,
integrated over the time variable produces a function singular as the two variables of the heat kernel approach to one another.
This singularity is the same as the singularity of the bound state wave function of a particle interacting with a delta source \cite{ermanturgut}, hence it is square integrable.

It is important to notice that $|\Psi_0 \rangle$ is not in the
domain of $H_0$. To prove this, it is sufficient to consider the
following term which appears in calculating $\langle \Psi_0 | H_0
| \Psi_0 \rangle$
\beqs
& & \hskip-0.5cm \int_{\mathcal{M}^2} \mathrm{d}^{2}_{g} x
\, \mathrm{d}^{2}_{g} y \; \int_{0}^{\infty} \mathrm{d} t_1 \;
e^{-t_1|E_{gr}|} K_{t_1}(x,y;g) \psi_0(y) \Bigg[ \int_{0}^{\infty}
\mathrm{d} t_2 \; e^{-t_2|E_{gr}|} \left(-{1\over 2m} \right)
\nabla_{g}^{2} K_{t_2}(x,y;g) \Bigg] \cr & & =
\int_{\mathcal{M}^2} \mathrm{d}^{2}_{g} x \, \mathrm{d}^{2}_{g} y
\; \int_{0}^{\infty} \mathrm{d} t_1 \; e^{-t_1|E_{gr}|}
K_{t_1}(x,y;g) \psi_0(y) \Bigg[ \int_{0}^{\infty} \mathrm{d} t_2
\; e^{-t_2|E_{gr}|} \bigg( -{\partial K_{t_2}(x,y;g) \over
\partial t_2} \bigg) \Bigg] \;,
\eeqs
\end{widetext}
where we have used the fact that the heat kernel satisfies the
heat equation (\ref{heatequation}).
After applying the integration by parts to the $t_2$ integral and
using the initial condition for the heat kernel $K_t(x,y;g)
\rightarrow \delta_g(x,y)$ as $t\rightarrow 0^{+}$ and
(\ref{semigroupprop}), we find
\begin{widetext}
\beqs
& & \int_{\mathcal{M}^2} \mathrm{d}^{2}_{g} x \,
\mathrm{d}^{2}_{g} y \; \int_{0}^{\infty} \mathrm{d} t_1 \;
e^{-t_1|E_{gr}|} K_{t_1}(x,y;g) \psi_0(y) \Bigg[ \delta_g(x,y) -
|E_{gr}| \int_{0}^{\infty} \mathrm{d} t_2 \; e^{-t_2|E_{gr}|}
K_{t_2}(x,y;g) \Bigg] \cr & & = \int_{\mathcal{M}}
\mathrm{d}^{2}_{g} x \; \int_{0}^{\infty} \mathrm{d} t_1 \;
e^{-t_1|E_{gr}|} K_{t_1}(x,x;g) \psi_0(x) \cr & & \hspace{4cm} -
|E_{gr}| \int_{0}^{\infty} \mathrm{d} t_1 \; e^{-t_1|E_{gr}|}
\int_{\mathcal{M}} \mathrm{d}^{2}_{g} x \; \int_{0}^{\infty}
\mathrm{d} t_2 \; e^{-t_2|E_{gr}|} K_{t_1+t_2}(x,x;g) \psi_0(x)
\;.
\eeqs
After the change of variables $u=t_1+t_2$ and $v=t_1-t_2$, we get
\beqs
& & \hskip-1cm \int_{\mathcal{M}} \mathrm{d}^{2}_{g} x \;
\psi_0(x) \Bigg[ \int_{0}^{\infty} \mathrm{d} t_1 \;
e^{-t_1|E_{gr}|} K_{t_1}(x,x;g) -  |E_{gr}|\int_{0}^{\infty}
\mathrm{d} u \; u \; e^{-u|E_{gr}|} K_{u}(x,x;g)\Bigg] \;.
\eeqs
\end{widetext}
The first term is divergent due to (\ref{asymheat}). Similar to
the problem with point interactions on manifolds which we studied
in \cite{ermanturgut}, our problem here can also be considered as
a kind of self-adjoint extension since the state $\Psi_0$ does not
belong to the domain of the free Hamiltonian. The self-adjoint
extension of the free Hamiltonian extends this domain such that
the state $\Psi_0$ is included. Although the state $\Psi_0$ is not
in the domain of $H_0$, the eigenvector corresponding to the
eigenfunction $u_0(x)$ for the lowest eigenvalue of $\Phi(E)$ can
be taken in the domain of $H_0$.

As a result,  $|\Psi_0 \rangle $ given in (\ref{psi1}) is not in
the product form in the large $n$ limit, that is,
\be
|\Psi_0 \rangle \neq \int_{\mathcal{M}^{n}} \mathrm{d}_{g}^{2}
y_1 \cdots \mathrm{d}_{g}^{2} y_n \; \prod_{k=1}^{n}
\Psi_{0}(y_{k}) |y_{1}\cdots y_{n} \rangle \;.
\ee
The solution takes a kind of  convolution of the wave functions in
the domain of $H_0$ with the bound state wave function which is
outside of this domain.

Yet, $\Phi(E)$'s lowest eigenfunction may well be approximated by
a product form for large number of bosons, that is,
\be
u_{0}(x_1,\cdots,x_{n-2})= u_0(x_1)\cdots u_0(x_{n-2}) \;,
\ee
with the normalization
\be
\label{normalization}
||u_0 ||^2 = \int_{\mathcal{M}} \mathrm{d}_{g}^{2} x \; |u_0(x)|^2
= 1 \;,\int_{\mathcal{M}} \mathrm{d}_{g}^{2} x \; |\psi_0(x)|^2 =1
\;.
\ee
Therefore, the expectation value of the principal operator by
applying the mean field ansatz must vanish, that is,
\beqs
\langle \phi_0 |\Phi(E_{gr})|\phi_0 \rangle = 0 \;.
\label{expvaluePhi}
\eeqs

Although such a mean field approximation is expected to be crude in less than three dimensions,
F. Calogero and A. Degasperis \cite{calogero} have shown that even in one dimension the mean field approach to this problem gives an
excellent agreement with the exact result. This
is a finite problem and we will see in the next subsection that the present approach is also consistent with the exact result.
Therefore, we expect that the mean field approximation to this problem in two-dimensions is also reliable.

In order to calculate (\ref{expvaluePhi}) explicitly, we will make normal
ordering of the creation and the annihilation operators by using
their eigenfunction expansion. Hence, the equation above yields
\begin{widetext}
\beqs
& & \int_{0}^{\infty} \mathrm{d} t \; \Bigg[ {e^{- t \mu^2 }
 \over 8 \pi t}
 - \int_{\mathcal{M}^2} \mathrm{d}_g^2 x \, \mathrm{d}_g^2 y \; \psi_{0}^{*}(x) \psi_{0}(y)
 K_{t}^{2}(x,y;g) e^{-t|E_{gr}|}
\bigg( \int_{\mathcal{M}^2} \mathrm{d}_g^2 x' \, \mathrm{d}_g^2 y'
\; u_{0}^{*}(x') K_t(x',y';g) u_0(y')\bigg)^{n-2}\Bigg] \cr & & =
{(n-2)(n-3) \over 2} \int_{0}^{\infty} \mathrm{d} t \;
\Bigg|\int_{\mathcal{M}^3} \mathrm{d}_g^2 x \,\mathrm{d}_g^2 x_1
\,\mathrm{d}_g^2 x_2 \; u_{0}^{*}(x_1) u_{0}^{*}(x_2)
K_{t}(x_1,x;g) \,
 K_{t}(x_2, x;g) \psi_0(x) \Bigg|^2 \cr &\ & \hspace{8cm} \times \; e^{-t |E_{gr}|}
\bigg( \int_{\mathcal{M}^2} \mathrm{d}_g^2 x' \, \mathrm{d}_g^2 y'
\; u_{0}^{*}(x') K_t(x',y';g) u_0(y')\bigg)^{n-4} \cr & & + \,
2(n-2) \int_{0}^{\infty} \mathrm{d} t \; \Bigg|
\int_{\mathcal{M}^2} \mathrm{d}_g^2 x \, \mathrm{d}_g^2 y \;
u_{0}^{*}(x) K_{t}(x,y;g) \psi_0(y) \Bigg|^2 e^{-t|E_{gr}|} \cr &
& \hspace{8cm} \times \; \bigg( \int_{\mathcal{M}^2}
\mathrm{d}_g^2 x' \, \mathrm{d}_g^2 y' \; u_{0}^{*}(x')
K_t(x',y';g) u_0(y')\bigg)^{n-3} \;. \label{meanfieldPhi}
\eeqs
\end{widetext}
We are going to approximately solve $E_{gr}$ from the above
equality for large values of $n$. In order to solve it, we may
assume that $|E_{gr}|$ grows rapidly with $n$. This is plausible
because $|E_{gr}| \simeq \mu^2 e^{\pi n/6}$ for flat space
$\mathbb{R}^2$ given in the mean field approximation
\cite{rajeevbound}. Every Riemannian manifold can locally be
considered as a flat space, and the infinity appears due to the
high values of momenta (ultraviolet divergence) or short distances
we expect that the result for the large $n$ behavior of the ground
state energy is similar on the manifold case. This allows us to
consider the above equality in the large values $|E_{gr}| \gg \mu^2$ so our
aim is to find only the terms that contribute most to the above
integrals.

We first calculate asymptotically the left hand side of
(\ref{meanfieldPhi})
\begin{widetext}
\beqs
& & \hskip-0.5cm \int_{0}^{\infty} \mathrm{d} t \; \Bigg[
{e^{- t \mu^2 }
 \over 8 \pi t}
 - \int_{\mathcal{M}^2} \mathrm{d}_g^2 x \, \mathrm{d}_g^2 y \; \psi_{0}^{*}(x) \psi_{0}(y)
 K_{t}^{2}(x,y;g) e^{-t|E_{gr}|}
\bigg( \int_{\mathcal{M}^2} \mathrm{d}_g^2 x' \, \mathrm{d}_g^2 y'
\; u_{0}^{*}(x') K_t(x',y';g) u_0(y')\bigg)^{n}\Bigg]  \;,
\label{meanfieldPhiLHS1}
\eeqs
\end{widetext}
for the large values of $|E_{gr}|$. We will now ignore the
additive constants to $n$, e.g., $n-2 \simeq n$ since $n$ is very
large. The major contribution to the above integral for large values of
$|E_{gr}|$ can be computed since the asymptotic
expansion of the following form, namely Laplace integrals
\be
I(|E_{gr}|)= \int_{a}^{b} \mathrm{d} t \; f(t) e^{-|E_{gr}|
g(t)} \;,
\ee
is given by Watson's Lemma \cite{bender}. The main contribution to
the above integral can be obtained by Taylor or when necessary by
the  asymptotic expansions of the functions $f(t)$ and $g(t)$ near
the minimum of $g(t)$. Similar to the reasoning given in the
previous section, we write the square of the heat kernel in a
subtle way, that is, we will use the initial condition for one of
the heat kernels near $t=0$. After this and an integration, we
substitute the asymptotic expansion (\ref{asymheat}) for the
diagonal heat kernel near $t=0$ (the region that gives the
dominant contribution). Hence, the left hand side for large values of
$|E_{gr}|\gg \mu^2$ becomes
\begin{widetext}
\beqs
& & \int_{0}^{\infty} \mathrm{d} t \; \Bigg[ {e^{- t \mu^2 }
 \over 8 \pi t}
 - \int_{\mathcal{M}} \mathrm{d}_g^2 x \; |\psi_{0}(x)|^2
 {e^{-t|E_{gr}|} \over 8 \pi t} \bigg(\int_{\mathcal{M}} \mathrm{d}_g^2 x'
\; |u_{0}(x')|^2 \bigg)^{n}\Bigg] \cr & & =  \int_{0}^{\infty}
\mathrm{d} t \; \Bigg[ {e^{- t \mu^2 }
 \over 8 \pi t}
 - {e^{-t|E_{gr}|} \over 8 \pi t}\Bigg] = {1 \over 8 \pi} \ln(|E_{gr}|/\mu^2) \;. \label{meanfieldPhiLHS2}
\eeqs
\end{widetext}

As for the right hand side of (\ref{meanfieldPhi}), we apply the
same method while we keep the next order terms coming from the
eigenfunction expansion of the heat kernel in the $n$-th power of
the integrals. Therefore, we obtain
\begin{widetext}
\beqs
& &
 {n^2 \over 2} \int_{0}^{\infty} \mathrm{d} t \;
\Bigg|\int_{\mathcal{M}} \mathrm{d}_g^2 x \; |u_{0}(x)|^{2}
\psi_0(x) \Bigg|^2   e^{-t|E_{gr}|} \bigg( \int_{\mathcal{M}}
\mathrm{d}_g^2 x \; |u_{0}(x)|^2
 - t K[u_0] \bigg)^{n} \cr & & \hspace{4cm} + \,
2n \int_{0}^{\infty} \mathrm{d} t \; \Bigg| \int_{\mathcal{M}}
\mathrm{d}_g^2 x \; u_{0}^{*}(x) \psi_0(x) \Bigg|^2 e^{-t|E_{gr}|}
\bigg( \int_{\mathcal{M}} \mathrm{d}_g^2 x \; |u_{0}(x)|^2 -  t
K[u_0] \bigg)^{n} \;, \label{meanfieldPhirhs 1st approximation}
\eeqs
\end{widetext}
where we have defined
\beqs
K[u_0]=\int_{\mathcal{M}} \mathrm{d}_g^2 x \; |\nabla_g
u_0(x)|^2 \;,
\eeqs
and used the eigenfunction expansion of the heat kernel
(\ref{heatkernelexpansion}) and expanded the exponential inside by
keeping the first two terms:
\be
K_t(x,y;g) \approx \sum_{l} \left(1-t \sigma_{l} \right)
f_l(x;g)f_l(y;g) \;.
\ee
We can rewrite the above expression (\ref{meanfieldPhirhs 1st
approximation}) by making a change of variable $t=t'/|E_{gr}|$ as
\begin{widetext}
\beqs
& &
 {n^2 \over 2 }\int_{0}^{\infty} {\mathrm{d} t' \over |E_{gr}| }\;
\Bigg|\int_{\mathcal{M}} \mathrm{d}_g^2 x \; |u_{0}(x)|^{2}
\psi_0(x) \Bigg|^2   e^{-t'} \Bigg[ \bigg( 1
 - {t' \over |E_{gr}|} K[u_0] \bigg)^{|E_{gr}|}\Bigg]^{n \over |E_{gr}|} \cr  & & \hspace{4cm} + \,
2n \int_{0}^{\infty} {\mathrm{d} t' \over |E_{gr}|} \; \Bigg|
\int_{\mathcal{M}} \mathrm{d}_g^2 x \; u_{0}^{*}(x) \psi_0(x)
\Bigg|^2 e^{-t'} \Bigg[\bigg(1 - {t' \over |E_{gr}|} K[u_0]
\bigg)^{|E_{gr}|} \Bigg]^{n \over |E_{gr}|} \;.
\label{meanfieldPhirhs 2nd approximation}
\eeqs
Moreover, we can think of terms in the square brackets as an
exponential for large values of $|E_{gr}|$ so that
\beqs
& &
 {n^2 \over 2} \int_{0}^{\infty} {\mathrm{d} t' \over |E_{gr}| }\;
\Bigg|\int_{\mathcal{M}} \mathrm{d}_g^2 x \; |u_{0}(x)|^{2}
\psi_0(x) \Bigg|^2   e^{-t'-{t' n \over |E_{gr}|} K[u_0]} \cr & &
\hspace{7cm} + 2n \int_{0}^{\infty} {\mathrm{d} t' \over |E_{gr}|}
\; \Bigg| \int_{\mathcal{M}} \mathrm{d}_g^2 x \; u_{0}^{*}(x)
\psi_0(x) \Bigg|^2 e^{-t'-{t' n\over |E_{gr}|} K[u_0]} \;.
\label{meanfieldPhirhs 3th approximation}
\eeqs
\end{widetext}
From equation (\ref{meanfieldPhi}), it is easy to see that the
left hand side is a monotonically increasing function and the
right hand side is a monotonically decreasing function of
$|E_{gr}|$ so there is a unique solution, say at $|E_{gr}|$.
Below this point $|E_{gr}|$, the left hand side is always less than the
right hand side. Therefore, if we can find an upper bound to the
right hand side of (\ref{meanfieldPhirhs 3th approximation}), and find a solution at $|E_*|$ this
implies that $E_{gr} \geq -|E_{*}|$. For this reason, let us
first set the normalized wave function of the orthofermion to saturate
the Cauchy-Schwarz inequality (as noted similarly in the flat case
\cite{rajeevbound})
\beqs
\psi_0(x) = {|u_0(x)|^2 \over \bigg(\int_{\mathcal{M}}
\mathrm{d}_g^2 x \; |u_0(x)|^4 \bigg)^{1/2}} \;.
\eeqs
Then, the upper bound of the right hand side of
(\ref{meanfieldPhirhs 3th approximation}) is
\begin{widetext}
\beqs
& &
 {n^2 \over 2
|E_{gr}|} {1 \over  (1+{n K[u_0] \over |E_{gr}|})}\;
\int_{\mathcal{M}} \mathrm{d}_g^2 x \; |u_{0}(x)|^{4} +  {2n \over
|E_{gr}|} {1 \over (1+{n K[u_0] \over |E_{gr}|})} {\bigg(
\int_{\mathcal{M}} \mathrm{d}_g^2 x \; u_{0}^{*}(x)|u_{0}(x)|^{2}
\bigg)^2 \over \int_{\mathcal{M}} \mathrm{d}_g^2 x \;
|u_{0}(x)|^{4}} \cr & & \hspace{4cm} \leq  {n^2 \over 2|E_{gr}|}
{1 \over (1+{n K[u_0] \over |E_{gr}|})}\; \int_{\mathcal{M}}
\mathrm{d}_g^2 x \; |u_{0}(x)|^{4} +  {2n \over |E_{gr}|} {1 \over
(1+{n K[u_0] \over |E_{gr}|})}  \;, \label{meanfieldPhirhs 4th
approximation}
\eeqs
\end{widetext}
where the Cauchy-Schwarz inequality in the second term is used,
that is,
\beqs
& & {\bigg(\int_{\mathcal{M}} \mathrm{d}_g^2 x \;
u_{0}^{*}(x)|u_{0}(x)|^{2} \bigg)^2 \over \int_{\mathcal{M}}
\mathrm{d}_g^2 x \; |u_{0}(x)|^{4}} \leq  {\bigg(||
|u_{0}(x)|^{2}|| \; ||u_{0}(x)|| \bigg)^2 \over \int_{\mathcal{M}}
\mathrm{d}_g^2 x \; |u_{0}(x)|^{4}} =1 \;.
\eeqs

We now recall the following theorem (Theorem 2.21 in
\cite{aubinbook}): The Sobolev imbedding theorem holds for a $D$
dimensional complete Riemannian manifold $\mathcal{M}$ with
bounded curvature and injectivity radius $\delta>0$. Moreover, for
any $\varepsilon>0$, there exists a constant $A_q(\varepsilon)$
such that every $\varphi \in H_{1}^{q}(\mathcal{M})$
($H_{1}^{q}(\mathcal{M})$ is the Sobolev space defined on a
manifold $\mathcal{M}$) satisfies
\beqs
||\varphi||_{p} \leq (\mathrm{K}(D,q) +
\varepsilon)||\nabla_g \varphi ||_q +
A_q(\varepsilon)||\varphi||_{q} \label{aubinbook} \;,
\eeqs
where $1/p= 1/q -1/D$ and
\begin{widetext}
\beqs
\mathrm{K}(D,q)={q-1 \over D-q}\bigg({D-q \over
D(q-1)}\bigg)^{1/q}\bigg({\Gamma(D+1) \over
\Gamma(D/q)\Gamma(D+1-D/q)\omega_{D-1}}\bigg)^{1/D} \;,
\eeqs
\end{widetext}
with $\omega_{D-1}$ is the volume of $\mathbb{S}_{D-1}$ of unit
radius.

Furthermore, there is an optimal inequality for the two
dimensional case given by T. Aubin \cite{aubinbook,aubin} and it
states that: Let $\mathcal{M}$ be a $D$ dimensional $C^{\infty}$
Riemannian manifolds with injectivity radius $\delta
>0$. If the curvature is constant or if the dimension is two and
the curvature is bounded, then $A_q(0)$ exists and every $\varphi
\in H_{1}^{q}(\mathcal{M})$ satisfies
\beqs
||\varphi||_{p} \leq \mathrm{K}(D,q) ||\nabla_g \varphi ||_q
+ A_q(0)||\varphi||_{q} \;. \label{aubin}
\eeqs
For $\mathbb{R}^{D}$ and $\mathbb{H}^D$, the inequality holds with
$A_q(0)=0$.

Let us choose $p=2$, $q=1$ and $D=2$ for our purposes, the
inequality (\ref{aubin}) is reduced to
\begin{widetext}
\beqs
\bigg(\int_{\mathcal{M}} \mathrm{d}_g^2 x \;
|\varphi(x)|^{2} \bigg)^{1/2} \leq  {2 \over \pi}
\int_{\mathcal{M}} \mathrm{d}_g^2 x \; |\nabla_g \varphi(x)| + A
\int_{\mathcal{M}} \mathrm{d}_g^2 x \; |\varphi(x)| \;,
\label{Aubin}
\eeqs
where $\mathrm{K}(2,1)=2/\pi$ and $A_1(0)=A$. If we set
$\varphi(x)=|u_{0}(x)|^{2}$, then
\beqs
\bigg( \int_{\mathcal{M}} \mathrm{d}_g^2 x \; |u_0(x)|^{4}
\bigg)^{1/2} & \leq & {2 \over \pi} \int_{\mathcal{M}}
\mathrm{d}_g^2 x \; |u_{0}^{*}(x) \nabla_g u_{0}(x)| + {2 \over
\pi} \int_{\mathcal{M}} \mathrm{d}_g^2 x \; |u_{0}(x) \nabla_g
u_{0}^{*}(x)| + \, A \int_{\mathcal{M}} \mathrm{d}_g^2 x \;
|u_0(x)|^2 \cr & \leq & A + {(4 / \pi)} K^{1/2}[u_0] \;,
\eeqs
where we have used Cauchy-Schwarz inequality and the normalization
of $u_0(x)$. Hence we obtain an upper bound for
(\ref{meanfieldPhirhs 4th approximation})
\beqs
 {n^2 \over 2|E_{gr}|} {\Bigg(A+ (4/\pi) K^{1/2}[u_0]\Bigg)^2 \over (1+{n
K[u_0] \over |E_{gr}|})}\;
 +  {2n \over
|E_{gr}|} {1 \over (1+{n K[u_0] \over |E_{gr}|})}  \;.
\label{meanfieldPhirhs 5th approximation}
\eeqs
\end{widetext}

Finally, combining the two results, we find that
\beqs
{|E_{gr}| \over 4 \pi} \ln(|E_{gr}|/\mu^2) \lesssim n^2 A^2
{ \bigg(1 + \beta z \bigg)^2 \over 1 + \alpha z^2} \;,
\eeqs
where $\alpha= 1/|E_{gr}|$, $\beta= 4 /(\pi A \sqrt{n})$ and
$z=\sqrt{n K[u_0]}$. For simplicity we ignore the second term in
the right hand side but we will return to these issues once we
find the solution and check the consistency of the approximations
that we have made so far. An upper bound of the right hand side is
achieved at $z_{*}=\beta/\alpha$ and its value is $n^2 A^2
(1+{\beta^2 \over \alpha})$. As a result of these, we eventually
obtain
\beqs
E_{gr} \gtrsim - \mu^2 e^{n(2^7/\pi)} \;.
\eeqs
We note that the location of this maximum for the variable $z$ is only formal, and does not correspond to the physical value of
$nK[u_0]$. It is simply chosen to get an upper bound for the right hand side, thus a lower bound for the energy. In fact, to be physically consistent,
$nK[u_0]$ should be of the order of $|E_{gr}|$ in the mean field approximation. Since we do not know a method to solve these equations,
it is not possible to calculate the actual values. Yet it is easy to check that in the limit where $nK[u_0] \gg |E_{gr}|$, the renormalized
term becomes dominated by this kinetic term, and the potential part also becomes much less than the renormalized term, hence there cannot be a zero
for the operator $\Phi(E)$ under these assumptions. Hence, we can keep $K[u_0] \ll |E_{gr}|$ condition in our approach. This has a nice interpretation
physically, for the $\Phi(E)$ operator, the ordinary total kinetic energy is of the order of the binding energy, moreover, the binding pair, transformed into
orthofermion, has  also finite kinetic energy. Nevertheless, we know that the actual wave function has infinite kinetic energy, thus this formalism nicely
takes out these pairs and converts them into regularly  interacting particles. As a result, they satisfy a nonlinear eigenvalue equation.

After we find the solution, it is easy to check the approximations
that we have made, the order of all these ignored terms are indeed
small. To be more precise, the next order terms coming from the
asymptotic expansion become lower order terms in $n$ for the
ground state energy.

\section{Confirmation of the Present Method in One Dimension}
\label{confirmation in one dimension}

We can apply our method to the ground state for the same
system in one dimension, where there is no need for
renormalization as can be easily seen from the short ``time"
asymptotic expansion of the heat kernel (\ref{asymheat}) in
(\ref{divergentterm}). The exact solution and the Hartree
approximation (for bosons) to the ground state in one dimension
have been studied in \cite{mcguire,calogero}. The exact solution
is given by \cite{mcguire}
\be
\Psi(x_1,\ldots,x_n)= C \exp\left(-{\lambda \over 4}
\sum_{i>j=1}^{n} |x_i-x_j|\right) \;, \label{exatGSWF}
\ee
where the normalization condition ($\int_{\mathbb{R}^n}
\mathrm{d}x_1 \ldots \mathrm{d}x_n \delta(x_{c.m}) |\Psi|^2 =n $)
allows us to calculate the constant $C$ explicitly \cite{mcguire}.
The exact ground state energy is then
\be
E_{gr}=-{\lambda^2 \over 48} n (n^2-1) \;. \label{exactGSE}
\ee
The Hartree solution to the ground state wave function (except for
the infinite degeneracy due to translational invariance) of the
same system \cite{calogero} is
\beqs
\Psi^{H}(x_1,\ldots,x_n) &=& n^{1/2} \psi(x_1)\cdots
\psi(x_n) \;, \cr  \psi(x) &=& {(\lambda n/8)^{1/2} \over \cosh
\left(\lambda n x /4 \right)} \;, \label{HartreeEF}
\eeqs
where $\int_{-\infty}^{\infty} \mathrm{d}x \; |\psi(x)|^2 =1$.
Since $n$ is large in this approximation, we may also write the
above solution as $(\lambda n/2)^{1/2}e^{-\lambda n |x|/4}$ and
the ground state energy is
\be
E_{gr}^{H}= - {\lambda^2 \over 48} n^2 (n-1) \;.
\label{HartreeGSE}
\ee
It is obvious that the exact results for the ground state
coincides with the results given in the Hartree approximation in
the large particle number limit.

Now, let us return to our method and calculate the principal
operator of the same system in $\mathbb{R}$, which is well defined
and finite from the beginning of the problem. The result is
\begin{widetext}
\beqs
&\ & \Phi(E)= {\Pi_1 \over \lambda} - \int_{\mathbb{R}^2}
\mathrm{d} x \, \mathrm{d} x' \chi^{\dag}(x)
 \int_{0}^{\infty} \mathrm{d} t \; K_{t}^{2}(x,x') e^{-t(H_0 -E)} \chi(x')
- {1 \over 2} \int_{\mathbb{R}^2} \mathrm{d} x \,\mathrm{d} x' \;
\chi^{\dag}(x)
 \bigg[ \int_{\mathbb{R}^4} \mathrm{d} x_1 \, \mathrm{d} x_2 \,
\mathrm{d} x'_1 \, \mathrm{d} x'_2 \cr & & \hspace{2cm} \times \;
\phi^{\dag}(x'_1)\,
 \phi^{\dag}(x'_2) \int_{0}^{\infty} \mathrm{d} t\; K_{t}(x'_1,x')
K_{t}(x',x'_2) \, K_{t}(x_1,x) \, K_{t}(x,x_2) \, \, e^{-t (H_0
-E)} \phi(x_1)\, \phi(x_2) \cr & & \hspace{2cm} + \; 4
\int_{\mathbb{R}^2} \mathrm{d} x_1 \, \mathrm{d} x_2 \;
\phi^{\dag}(x_1) \int_{0}^{\infty} \mathrm{d} t \; K_{t}(x_1,x')
\,K_{t}(x',x) \, K_{t}(x,x_2) \, e^{-t (H_0 -E)} \phi(x_2)
 \bigg] \chi(x') \;, \label{Phi1D}
\eeqs
where $K_t(x,y)= {e^{-|x-y|^2/4t} \over (4\pi t)^{1/2}}$. The
condition (\ref{expvaluePhi}) gives
\beqs
& & {1 \over \lambda} - \int_{0}^{\infty} \mathrm{d} t \;
 \int_{\mathbb{R}^2} \mathrm{d} x \, \mathrm{d} y \; \psi_{0}^{*}(x) \psi_{0}(y)
 K_{t}^{2}(x,y) e^{-t|E_{gr}|}
\bigg( \int_{\mathbb{R}^2} \mathrm{d} x' \, \mathrm{d} y' \;
u_{0}^{*}(x') K_t(x',y') u_0(y')\bigg)^{n-2} \cr & & =
{(n-2)(n-3)\over 2}\int_{0}^{\infty} \mathrm{d} t \;
\Bigg|\int_{\mathbb{R}^3} \mathrm{d} x \,\mathrm{d} x_1
\,\mathrm{d} x_2 \; u_{0}^{*}(x_1) u_{0}^{*}(x_2) K_{t}(x_1,x) \,
 K_{t}(x_2, x) \psi_0(x) \Bigg|^2 \cr &\ & \hspace{8cm} \times \; e^{-t |E_{gr}|}
\bigg( \int_{\mathbb{R}^2} \mathrm{d} x' \, \mathrm{d} y' \;
u_{0}^{*}(x') K_t(x',y') u_0(y')\bigg)^{n-4} \cr & & + \; 2(n-2)
\int_{0}^{\infty} \mathrm{d} t \; \Bigg| \int_{\mathbb{R}^2}
\mathrm{d} x \, \mathrm{d} y \; u_{0}^{*}(x) K_{t}(x,y) \psi_0(y)
\Bigg|^2 e^{-t|E_{gr}|} \bigg( \int_{\mathbb{R}^2} \mathrm{d} x'
\, \mathrm{d} y' \; u_{0}^{*}(x') K_t(x',y') u_0(y')\bigg)^{n-3}
\;. \label{meanfieldPhi1D}
\eeqs
\end{widetext}
Following the same analysis given above, we find the left hand
side of (\ref{meanfieldPhi1D}) for large values of $|E_{gr}|$
\beqs
& & {1 \over \lambda}
 - \int_{\mathbb{R}} \mathrm{d} x \; |\psi_{0}(x)|^2
 {e^{-t|E_{gr}|} \over (8 \pi t)^{1/2}} \bigg(\int_{\mathbb{R}} \mathrm{d} x'
\; |u_{0}(x')|^2 \bigg)^{n} = {1 \over \lambda} - {1 \over 2
\sqrt{2 |E_{gr}|}} \;, \label{meanfieldPhiLHS21D}
\eeqs
and the right hand side of it in the same limit, which is the
analog of (\ref{meanfieldPhirhs 4th approximation}) in one
dimension, becomes less than the following term
\beqs
{n^2 \over 2 |E_{gr}|} {1 \over  (1+{n K[u_0] \over
|E_{gr}|})}\; \int_{\mathbb{R}} \mathrm{d} x \; |u_{0}(x)|^{4} +
{2n \over |E_{gr}|} {1 \over (1+{n K[u_0] \over |E_{gr}|})} \;.
\label{meanfieldPhirhs 4th approximation1D}
\eeqs
In one dimension, the Sobolev inequality for $2<q<\infty$ is given
as \cite{Frank}
\beqs
\bigg(\int_{\mathbb{R}} \mathrm{d} x \; \left|{\mathrm{d}
u_0 \over \mathrm{d} x}\right|^2 \bigg)^\theta \bigg(
\int_{\mathbb{R}} \mathrm{d} x \; |u_0|^2 \bigg)^{1-\theta} \geq
S_{1,q} \bigg( \int_{\mathbb{R}} \mathrm{d} x \; |u_0|^q
\bigg)^{2/q}\;, \label{sobolev1D}
\eeqs
where $\theta= {1 \over 2}\left(1-{2 \over q}\right)$ and
\beqs
S_{1,q}= {q \theta^{\theta} (1-\theta)^{1-\theta} \over
2^{2/q} (q-2)^{(q-2)/q}} \left[{\sqrt{\pi} \Gamma\left({q \over
q-2}\right) \over \Gamma \left({q \over q-2}+ {1 \over
2}\right)}\right]^{(q-2)/q} \label{S}
\eeqs
with equality if and only if $u_0(x)=c \cosh^{-2/(q-2)}(b(x-a))$
for some $a \in \mathbb{R}$, $b>0$ and $c \in \mathbb{C}$. Since
we are looking for an upper bound to (\ref{meanfieldPhirhs 4th
approximation1D}) we will choose $q=4$ so that $\theta=1/4$. Then
the Sobolev inequality in  (\ref{sobolev1D}) gives
\beqs
\int_{\mathbb{R}} \mathrm{d} x \; |u_0|^4 \leq S_{1,4}^{-2}
\bigg( \int_{\mathbb{R}} \mathrm{d} x \; \left|{\mathrm{d} u_0
\over \mathrm{d} x}\right|^2 \bigg)^{1/2} \bigg( \int_{\mathbb{R}}
\mathrm{d} x \; |u_0|^2 \bigg)^{3/2} = {1 \over \sqrt{3}}
K^{1/2}[u_0] \;, \label{sobolev1Dq=4}
\eeqs
where we have used the normalization of the wave functions and
$S_{1,4}=3^{1/4}$. Using this result in (\ref{meanfieldPhirhs 4th
approximation1D}) and from (\ref{meanfieldPhiLHS21D}), we get
\beqs
& & {1 \over \lambda} - {1 \over 2 \sqrt{2 |E_{gr}|}} \leq
{n^2 \over 2 \sqrt{3}|E_{gr}|} {K^{1/2}[u_0] \over  (1+{n K[u_0]
\over |E_{gr}|})} + {2n \over |E_{gr}|} {1 \over (1+{n K[u_0]
\over |E_{gr}|})} \;.
\eeqs
Keeping the leading order term on both sides, we obtain
\be
{1 \over \lambda}  \leq {n^2 \over 2 \sqrt{3}|E_{gr}|}
{K^{1/2}[u_0] \over  (1+{n K[u_0] \over |E_{gr}|})} \;.
\ee
Let us define the variables $z=n K[u_0]$  and $\alpha=1/|E_{gr}|$,
and then find the upper bound to the right hand side. This occurs
at $z=1/\alpha$ so we get
\be
E_{gr} \geq - {\lambda^2 \over 48} n^3 \;,
\ee
which is exactly the same result given in (\ref{HartreeGSE}) in
the leading order. We note that in this approach the kinetic
energy of the center of mass motion is automatically set to be
zero. We can also find the eigenfunction from our analysis. As a
result of the above theorem, the Sobolev inequality that we have
used above is saturated if
\be
u_0(x)= {\sqrt{b/2} \over \cosh(b x)} \;. \label{saturated}
\ee
Here we have chosen the constant $a=0$ without loss of generality
and the coefficient $c=\sqrt{b/2}$ has been found from the
normalization. The constant $b$ can be determined from the
solution $z= n K[u_0]= |E_{gr}|$. Since the saturating solution
(\ref{saturated}) satisfies
\be
\int_{\mathbb{R}} \mathrm{d} x \; |u_0|^4 = {1 \over \sqrt{3}}
K^{1/2}[u_0] \;,
\ee
we obtain $b=\lambda n/4$. Therefore we find exactly the same
result obtained from the Hartree approximation (\ref{HartreeEF}).
Incidentally, in this limit the wave functions could be taken as,
\be
u_0(x)=\sqrt{{ \lambda n \over 2}} e^{-n\lambda |x|/4} \;.
\ee
and they are related to the actual wave function of the system by
our previous formula (\ref{psi0}).

\section{Renormalization Group Equations}
\label{Renormalization Group}

The renormalization group equations (or Callan-Symanzik equations)
for the system, where the particles do not interact with each
other but interact with an external Dirac delta potential in two
and three dimensional flat spaces, has been worked out in
\cite{adhikari,Camblong2,manuel2}. Many-body version of the same
problem, where the particles interact via two-body delta
potentials, has also been studied \cite{bergman1, Bergman,
bergman2}.

Recently, we have derived the generalization of the
renormalization group equations of the above one-body model with
$N$ delta centers into two and three dimensional Riemannian
manifolds \cite{ermanturgut}. Here, we will show that the
interacting version of the problem can be also studied explicitly,
as we will see.

One possible way for the renormalization scheme in order to
determine how the coupling constant changes with the energy scale
is to define the following renormalized coupling constant
$\lambda_{R}(M)$ in terms of the bare coupling constant
$\lambda(\epsilon)$
\be
{1 \over \lambda_{R}(M)} = {1 \over \lambda(\epsilon)} -
\int_{\epsilon}^{\infty} \mathrm{d} t {e^{- M^{2} t} \over 8 \pi
t} \;,  \label{renormalized coupling}
\ee
where $M$ is the renormalization scale (it is of dimension
$[E]^{1/2}$). Then, the renormalized principal operator in terms
of renormalized coupling constant is given by
\begin{widetext}
\beqs
&\ & \Phi^{R}(E)= {\Pi_1 \over \lambda_{R}(M)} -
\int_{\mathcal{M}^2} \mathrm{d}_g^2 x \, \mathrm{d}_g^2 x'
\chi^{\dag}_{g}(x)
 \int_{0}^{\infty} \mathrm{d} t \; \bigg[ K_{t}^{2}(x,x';g) e^{-t(H_0 -E)} - {e^{- t M^2 }
 \over 8 \pi t}
\delta_g^{(2)}(x,x') \bigg] \chi_{g}(x')
  \cr & & - {1\over 2}
\int_{\mathcal{M}^2} \mathrm{d}_g^2 x \,\mathrm{d}_g^2 x' \;
\chi^{\dag}_{g}(x)
 \bigg[ \int_{\mathcal{M}^4} \mathrm{d}_g^2 x_1 \, \mathrm{d}_g^2 x_2 \,
\mathrm{d}_g^2 x'_1 \, \mathrm{d}_g^2 x'_2 \;
\phi^{\dag}_{g}(x'_1)\,
 \phi^{\dag}_{g}(x'_2) \int_{0}^{\infty} \mathrm{d} t\; K_{t}(x_1,x;g) \,
 K_{t}(x_2, x;g) \cr &\ & \times \, K_{t}(x',x'_1;g) \,
 K_{t}(x',x'_2;g)\, e^{-t (H_0 -E)}
\phi_{g}(x_1)\, \phi_{g}(x_2) + 4 \int_{\mathcal{M}^2}
\mathrm{d}_g^2 x_1 \, \mathrm{d}_g^2 x_2 \; \phi^{\dag}_{g}(x_1) \
\cr & & \hspace{4cm} \times \, \int_{0}^{\infty} \mathrm{d} t
\;K_{t}(x_2,x;g) \,
 K_{t}(x',x;g)\, K_{t}(x',x_1;g)\, e^{-t (H_0 -E)} \phi_{g}(x_2)
 \bigg] \chi_{g}(x') \;\textcolor[rgb]{1.00,0.00,0.00}{.} \label{RenormalizedPhi}
\eeqs
\end{widetext}

Here the bound state energies  are  again determined from the condition $
\Phi^{R} (E)|\Psi \rangle =0$ in the $n$-particle sector, however there is an ambiguity, we have a family of solutions
for different choices of $M$ and $\lambda_R(M)$. To determine the value   of $\lambda_{R}(M)$ at an arbitrary
value of the renormalization point $M$, a natural choice would be to use the
physically measured two-body bound state energy $E^{(2)}_{gr}$, if it exits, otherwise to use a scattering amplitude at some two particle energy.
The solution then determines the
relation between $\lambda_{R}(M)$ and $M$. Explicit dependence on
$M$ cancels the implicit dependence on $M$ through
$\lambda_{R}(M)$.
In the case of two-body bound state energy, the principal operator acts on $|0\rangle
\otimes \int_{\mathcal{M}} \mathrm{d}_{g}^{2}x \; \psi(x) \chi^{\dag}_{g}|0 \rangle$. Hence, because of the condition for the bound
states (\ref{spectrum}) we obtain an equation, the solution of which fixes $\lambda_R(M)$ as a function of $M, E_{gr}^{(2)}$:
\begin{widetext}
\beqs
& &  {1 \over \lambda_{R}(M)} - \int_{\mathcal{M}^2}
\mathrm{d}_g^2 x \, \mathrm{d}_g^2 x' \psi^{*}(x)
 \int_{0}^{\infty} \mathrm{d} t \; \bigg[ K_{t}^{2}(x,x';g) e^{-t |E_*|} - {e^{- t M^2 }
 \over 8 \pi t}
\delta_g^{(2)}(x,x') \bigg] \psi(x')  = 0 \;.
\eeqs
\end{widetext}
Even if we cannot explicitly
solve this equation, the arbitrariness in the choice of the scale is reflected by expression below,
\be
M {\mathrm{d} \Phi^{R}(M,\lambda_R(M),E;g) \over \mathrm{d} M}
=0 \;,\label{renorm cond}
\ee
or
\be
\left( M {\partial \over \partial M} + \beta(\lambda_R)
{\partial \over \partial \lambda_R}
\right)\Phi^{R}(M,\lambda_R(M),E;g)=0 \;,\label{renorm condition
2d}
\ee
where
\be
\label{beta function def} \beta(\lambda_R)= M {\partial
\lambda_R \over
\partial M}
\ee
is called the $\beta$ function and equation (\ref{renorm condition
2d}) is the renormalization group (RG) equation. This equation implies that the physics is independent of the choice of our renormalization scale. Using
(\ref{RenormalizedPhi}) in (\ref{renorm condition 2d}), we can
find $\beta$ function exactly
\be
\beta(\lambda_R) = -{\lambda_{R}^{2} \over 4 \pi} < 0 \;.
\ee
This result is exactly the same as the one in flat spaces given in
the literature \cite{Bergman} so our problem is asymptotically
free, too.

We will now derive an analog of Callan-Symanzik equation for our principal operator $\Phi_R$ and show that there is a simple solution of this
equation, related to the flow of the renormalized coupling constant. This will reconcile present method with the tools of conventional approach to field theories.

In order to see this, we will use scaling property of the heat kernel in two
dimensional Riemannian manifolds
\be
K_{t}(x,y;g)=\gamma^{-2} K_{\gamma^{-2} t}(x,y;\gamma^{-2} g)
\;,\label{heatkernelscaling2}
\ee
with the assumption that the manifold that we are interested in is
stochastically complete, that is, $\int_{\mathcal{M}}
\mathrm{d}^{2}_{g}x \; K_t(x,y;g)=1$.  There exists a unitary
representation for the scaling transformation of the metric $g
\mapsto \gamma^{-2}g$ such that the creation and annihilation
operators transform like
\begin{widetext}
\beqs
U(\gamma) \phi_g(x) U^{\dagger}(\gamma)& = &
 \gamma^{-1} \phi_{\gamma^{-2}g}(x) \;, \hspace {1cm} U(\gamma) \phi_{g}^{\dagger}(x)
 U^{\dagger}(\gamma) =
 \gamma^{-1} \phi_{\gamma^{-2}g}^{\dagger}(x)
\cr  U(\gamma) \chi_g(x) U^{\dagger}(\gamma) &=& \gamma^{-1}
\chi_{\gamma^{-2}g}(x) \;, \hspace{1cm} U(\gamma)
\chi_{g}^{\dagger}(x) U^{\dagger}(\gamma) = \gamma^{-1}
\chi_{\gamma^{-2}g}^{\dagger}(x) \;, \label{transformedfields}
\eeqs
\end{widetext}
where we have used their commutation relations and the algebra of
the orthofermions defined in (\ref{orthofermion algebra}). Wave function
normalization will be invariant under this transformation.

Let us first simultaneously scale the energy by $\gamma^2$ and the
metric by $\gamma^{-2}$ in the renormalized principal operator
given explicitly in (\ref{RenormalizedPhi}) and get
\begin{widetext}
\beqs
& & \hskip-2cm \Phi^{R}(M, \lambda_R(M), \gamma^2
E;\gamma^{-2}g) = {\int_{\mathcal{M}}
\mathrm{d}^{2}_{\gamma^{-2}g} x \;
\chi_{\gamma^{-2}g}^{\dagger}(x) \chi_{\gamma^{-2}g}(x) \over
\lambda_{R}(M)} - \; \int_{\mathcal{M}^2}
\mathrm{d}_{\gamma^{-2}g}^2 x \, \mathrm{d}_{\gamma^{-2}g}^2 x'
\chi^{\dag}_{\gamma^{-2}g}(x)
 \cr & & \times \; \int_{0}^{\infty} \mathrm{d} t \; \bigg[ K_{t}^{2}(x,x';\gamma^{-2}g) e^{-t(H_0 -\gamma^2 E)} - {e^{- t M^2 }
 \over 8 \pi t}
\delta_{\gamma^{-2} g}^{(2)}(x,x') \bigg] \chi_{\gamma^{-2} g}(x')
  \cr & & - \; {1\over 2}
\int_{\mathcal{M}^2} \mathrm{d}_{\gamma^{-2} g}^2 x
\,\mathrm{d}_{\gamma^{-2} g}^2 x' \; \chi^{\dag}_{\gamma^{-2}
g}(x)
 \bigg[ \int_{\mathcal{M}^4} \mathrm{d}_{\gamma^{-2} g}^2 x_1 \, \mathrm{d}_{\gamma^{-2} g}^2 x_2 \,
\mathrm{d}_{\gamma^{-2} g}^2 x'_1 \, \mathrm{d}_{\gamma^{-2} g}^2
x'_2  \phi^{\dag}_{\gamma^{-2} g}(x'_1)\,
  \cr & & \times \; \phi^{\dag}_{\gamma^{-2} g}(x'_2) \int_{0}^{\infty} \mathrm{d} t\; K_{t}(x_1,x;\gamma^{-2} g) \,
 K_{t}(x_2, x;\gamma^{-2} g) K_{t}(x',x'_1; \gamma^{-2} g) \,
 K_{t}(x',x'_2;\gamma^{-2} g) \cr & & \times \; e^{-t (H_0 -\gamma^2 E)}
\phi_{\gamma^{-2} g}(x_1)\, \phi_{\gamma^{-2} g}(x_2) + 4
\int_{\mathcal{M}^2} \mathrm{d}_{\gamma^{-2} g}^2 x_1 \,
\mathrm{d}_{\gamma^{-2} g}^2 x_2 \; \phi^{\dag}_{\gamma^{-2}
g}(x_1) \ \cr &\ & \times \int_{0}^{\infty} \mathrm{d} t
\;K_{t}(x_2,x;\gamma^{-2} g) \,
 K_{t}(x',x; \gamma^{-2} g) K_{t}(x',x_1; \gamma^{-2} g)  e^{-t (H_0 -\gamma^2 E)} \phi_{\gamma^{-2} g}(x_2)
 \bigg] \chi_{\gamma^{-2} g}(x') \;. \label{scaledPhi1}
\eeqs
Now we make a change of variable $t\mapsto \gamma^{-2} t$ and use
the scaling property of the heat kernel (\ref{heatkernelscaling2})
and obtain
\beqs
& & \hskip-1.8cm \Phi^{R}(M, \lambda_R(M), \gamma^2
E;\gamma^{-2}g) = {\gamma^{-2} \int_{\mathcal{M}}
\mathrm{d}^{2}_{g} x \; \chi_{\gamma^{-2}g}^{\dagger}(x)
\chi_{\gamma^{-2}g}(x) \over \lambda_{R}(M)} -
\int_{\mathcal{M}^2} \mathrm{d}_{g}^2 x \, \mathrm{d}_{g}^2 x'
\chi^{\dag}_{\gamma^{-2}g}(x) \cr & & \hspace{3cm} \times
 \int_{0}^{\infty} \mathrm{d} t \; \gamma^{-2}
 \bigg[K_{t}^{2}(x,x';g) e^{-t\gamma^{-2}(H_0 -\gamma^2 E)} - {e^{- t \gamma^{-2} M^2 }
 \over 8 \pi t}
\delta_{g}^{(2)}(x,x') \bigg] \chi_{\gamma^{-2} g}(x')
  \cr & & - {1\over 2}
\int_{\mathcal{M}^2} \mathrm{d}_{g}^2 x \,\mathrm{d}_{g}^2 x' \;
\chi^{\dag}_{\gamma^{-2} g}(x)
 \bigg[ \gamma^{-6} \int_{\mathcal{M}^4} \mathrm{d}_{g}^2 x_1 \, \mathrm{d}_{g}^2 x_2 \,
\mathrm{d}_{g}^2 x'_1 \, \mathrm{d}_{g}^2 x'_2
\phi^{\dag}_{\gamma^{-2} g}(x'_1)\,
  \cr & & \hspace{3cm} \times \; \phi^{\dag}_{\gamma^{-2} g}(x'_2) \int_{0}^{\infty} \mathrm{d} t\; K_{t}(x_1,x;g) \,
 K_{t}(x_2, x;g) K_{t}(x',x'_1; g) \,
 K_{t}(x',x'_2;g) \cr & & \times \; e^{-t \gamma^{-2} (H_0 -\gamma^2 E)}
\phi_{\gamma^{-2} g}(x_1)\, \phi_{\gamma^{-2} g}(x_2) + 4
\gamma^{-4} \int_{\mathcal{M}^2} \mathrm{d}_{g}^2 x_1 \,
\mathrm{d}_{g}^2 x_2 \; \phi^{\dag}_{\gamma^{-2} g}(x_1) \ \cr &\
& \hspace{2cm} \times \; \int_{0}^{\infty} \mathrm{d} t
\;K_{t}(x_2,x;g) \,
 K_{t}(x',x; g) K_{t}(x',x_1; g) e^{-t \gamma^{-2} (H_0 -\gamma^2 E)}
\phi_{\gamma^{-2} g}(x_2)
 \bigg] \chi_{\gamma^{-2} g}(x') \;, \label{scaledPhi2}
\eeqs
\end{widetext}
where we have used $\delta_{\gamma^{-2}g}^{(2)}(x,x')= \gamma^2
\delta_{g}^{(2)}(x,x')$ and $\mathrm{d}^{2}_{\gamma^{-2}g}x =
\gamma^{-2} \mathrm{d}^{2}_{g}x $. Using
(\ref{transformedfields}), and inserting the identity
$U(\gamma)U^{\dagger}(\gamma)$ in the appropriate places inside
the above equation, we obtain for $U^{\dagger}(\gamma) \;
\Phi^{R}(M, \lambda_R(M), \gamma^2 E;\gamma^{-2}g) \; U(\gamma)$:
\begin{widetext}
\beqs
& & U^{\dagger}(\gamma) \; \Phi^{R}(M, \lambda_R(M),
\gamma^2 E;\gamma^{-2}g) \; U(\gamma) = {\Pi_1 \over
\lambda_{R}(M)} - \int_{\mathcal{M}^2} \mathrm{d}_{g}^2 x \,
\mathrm{d}_{g}^2 x' \chi^{\dag}_{g}(x) \cr & & \hspace{5cm} \times
 \int_{0}^{\infty} \mathrm{d} t \; \bigg[ K_{t}^{2}(x,x';g) e^{-t(H_0 -E)} - {e^{- t(\gamma^{-1} M)^2 }
 \over 8 \pi t}
\delta_{g}^{(2)}(x,x') \bigg] \chi_{g}(x')
  \cr & & - {1\over 2}
\int_{\mathcal{M}^2} \mathrm{d}_{g}^2 x \,\mathrm{d}_{g}^2 x' \;
\chi^{\dag}_{g}(x)
 \bigg[\int_{\mathcal{M}^4} \mathrm{d}_{g}^2 x_1 \, \mathrm{d}_{g}^2 x_2 \,
\mathrm{d}_{g}^2 x'_1 \, \mathrm{d}_{g}^2 x'_2
\phi^{\dag}_{g}(x'_1)\, \phi^{\dag}_{g}(x'_2) \int_{0}^{\infty}
\mathrm{d} t\; K_{t}(x_1,x;g) \,
 K_{t}(x_2, x;g) \cr & & \hspace{5cm} \times \; K_{t}(x',x'_1; g) \,
 K_{t}(x',x'_2;g)  e^{-t(H_0 -E)}
\phi_{g}(x_1)\, \phi_{g}(x_2) + 4 \int_{\mathcal{M}^2}
\mathrm{d}_{g}^2 x_1 \, \mathrm{d}_{g}^2 x_2 \;
\phi^{\dag}_{g}(x_1) \cr & & \hspace{4cm} \times \;
\int_{0}^{\infty} \mathrm{d} t \;K_{t}(x_2,x;g) \,
 K_{t}(x',x; g)\, K_{t}(x',x_1; g)  e^{-t (H_0 - E)} \phi_{g}(x_2)
 \bigg] \chi_{g}(x') \;, \label{scaledPhi3}
\eeqs
\end{widetext}
where
\be
U^{\dagger}(\gamma) e^{-t\gamma^{-2}(H_0-\gamma^2 E)}
U(\gamma)=  e^{-t(H_0-E)} \;.
\ee
Therefore we finally obtain
\beqs
& & U^{\dagger}(\gamma) \; \Phi^{R}(M, \lambda_R(M),
\gamma^2 E;\gamma^{-2}g) \; U(\gamma)=
\Phi^{R}(\gamma^{-1} M, \lambda_R(M), E;g) \;.
\label{transformedPhi}
\eeqs
It is important to note that we need to scale the metric as well.
The idea of the metric scaling in deriving the renormalization
group equation was motivated by \cite{odintsov} in the context of
renormalization group in quantum field theory on curved spaces.
Hence we have
\beqs
& & \gamma {\mathrm{d} \over \mathrm{d} \gamma}
\bigg[U^{\dagger}(\gamma) \Phi^{R}(M, \lambda_R(M), \gamma^2
E;\gamma^{-2}g) U(\gamma) =
\Phi^{R}(\gamma^{-1} M, \lambda_R(M), E; g) \bigg] \;.
\eeqs
This leads to the renormalization group equation for
$U^{\dagger}(\gamma) \Phi^{R}(M, \lambda_R(M), \gamma^2
E;\gamma^{-2}g) U(\gamma)$
\begin{widetext}
\beqs
& & \gamma {\mathrm{d} \over \mathrm{d} \gamma}
U^{\dagger}(\gamma) \Phi^{R}(M, \lambda_R(M), \gamma^2
E;\gamma^{-2}g) U(\gamma) + \, M {\partial \over
\partial M} U^{\dagger}(\gamma) \Phi^{R}(M, \lambda_R(M), \gamma^2
E;\gamma^{-2}g) U(\gamma)=0 \;,
\eeqs
or
\be
\left[\gamma {\mathrm{d} \over \mathrm{d} \gamma} -
\beta(\lambda_R) {\partial \over
\partial \lambda_R}\right] U^{\dagger}(\gamma)\Phi^{R}(M, \lambda_R(M), \gamma^2
E;\gamma^{-2}g) U(\gamma)=0 \;. \label{rg equation 2d}
\ee
\end{widetext}
If we postulate the following functional form for the principal
matrix
\beqs
& & U^{\dagger}(\gamma) \Phi^{R}(M, \lambda_R(M), \gamma^2
E;\gamma^{-2}g) U(\gamma) = f(\gamma)
\Phi^{R}(M, \lambda_R(\gamma M), E;g) \label{funtional ansatz}
\;,
\eeqs
and substitute into (\ref{rg equation 2d}) we obtain an ordinary
differential equation for the function $f$
\be
\gamma {\mathrm{d} f(\gamma) \over \mathrm{d} \gamma} =0 \;.
\ee
This has the solution $f(\gamma)=1$ using the initial condition
at $\gamma=1$. Therefore, we get
\beqs
& & U^{\dagger}(\gamma) \Phi^{R}(M, \lambda_R(M), \gamma^2
E;\gamma^{-2}g) U(\gamma) = \Phi^{R}(M,
\lambda_R(\gamma M), E;g) \;, \label{grm scaling 2d}
\eeqs
which means that there is no anomalous scaling. This interesting result has been derived in \cite{bergman1, Bergman}
for the two-particle sector in flat space for $T$-matrix.

By integrating
\be
\beta(\lambda_R) = \bar{M} {\partial \lambda_R(\bar{M})\over
\partial \bar{M}} = -{\lambda_{R}^{2}(\bar{M}) \over 4 \pi}
\ee
between  $\bar{M}= M$ to $\bar{M}=\gamma M$ we can find the flow
equation for the coupling constant
\be
\lambda_R(\gamma M) = {\lambda_R(M) \over 1+ {1 \over 4 \pi}
\lambda_R(M)
 \ln \gamma} \;. \label{coupling const evolve 2d}
\ee
Indeed, the above evolution can also be derived from the choice of our coupling constant given in (\ref{renormalized coupling}). 
One can explicitly check the relation (\ref{grm scaling 2d}) if
the coupling constant evolves according to (\ref{coupling const
evolve 2d}). First, we add and subtract a term in the time
integral to $\Phi^{R}(M, \lambda_R(\gamma M), E;g)$ (as indicated
explicitly below) and use (\ref{coupling const evolve 2d}):
\begin{widetext}
\beqs
& & \Phi^{R}(M, \lambda_R(\gamma M), E;g) = {\Pi_1 \over
\lambda_{R}(M)} + {\Pi_1 \over 4 \pi} \ln \gamma -
\int_{\mathcal{M}^2} \mathrm{d}_g^2 x \, \mathrm{d}_g^2 x'
\chi^{\dag}_{g}(x) \cr & & \hspace{1cm} \times \;
 \int_{0}^{\infty} \mathrm{d} t \; \bigg[ K_{t}^{2}(x,x';g) e^{-t(H_0 -E)} - {e^{- t M^2 }
 \over 8 \pi t}
\delta_g^{(2)}(x,x') + {e^{- t \gamma^{-2}M^2 }
 \over 8 \pi t}
\delta_g^{(2)}(x,x') -{e^{- t \gamma^{-2} M^2 }
 \over 8 \pi t}
\delta_g^2(x,x') \bigg] \chi_{g}(x') \cr & & \hspace{1cm}  -
{1\over 2} \int_{\mathcal{M}^2} \mathrm{d}_g^2 x \,\mathrm{d}_g^2
x' \; \chi^{\dag}_{g}(x)  \bigg[ \int_{\mathcal{M}^4}
\mathrm{d}_g^2 x_1 \, \mathrm{d}_g^2 x_2 \, \mathrm{d}_g^2 x'_1 \,
\mathrm{d}_g^2 x'_2 \; \phi^{\dag}_{g}(x'_1)\,
 \phi^{\dag}_{g}(x'_2) \int_{0}^{\infty} \mathrm{d} t\; K_{t}(x_1,x;g) \,
 K_{t}(x_2, x;g) \cr &\ & \hspace{3cm} \times \; K_{t}(x',x'_1;g) \,
 K_{t}(x',x'_2;g)\, e^{-t (H_0 -E)}
\phi_{g}(x_1)\, \phi_{g}(x_2) + 4 \int_{\mathcal{M}^2}
\mathrm{d}_g^2 x_1 \, \mathrm{d}_g^2 x_2 \; \phi^{\dag}_{g}(x_1) \
\cr &\ & \hspace{4cm} \times \; \int_{0}^{\infty} \mathrm{d} t
\;K_{t}(x_2,x;g) \,
 K_{t}(x',x;g)\, K_{t}(x',x_1;g)\, e^{-t (H_0 -E)} \phi_{g}(x_2)
 \bigg] \chi_{g}(x') \;.
\eeqs
we find
\beqs
& & \hskip-2cm \Phi^{R}(M, \lambda_R(\gamma M), E;g) =
{\Pi_1 \over \lambda_{R}(M)} - \int_{\mathcal{M}^2} \mathrm{d}_g^2
x \, \mathrm{d}_g^2 x' \chi^{\dag}_{g}(x) \cr & & \times \;
 \int_{0}^{\infty} \mathrm{d} t \; \bigg[ K_{t}^{2}(x,x';g) e^{-t(H_0 -E)} -{e^{- t \gamma^{-2} M^2 }
 \over 8 \pi t}
\delta_g^{(2)}(x,x') \bigg] \chi_{g}(x')
 - {1\over 2}
\int_{\mathcal{M}^2} \mathrm{d}_g^2 x \,\mathrm{d}_g^2 x' \;
\chi^{\dag}_{g}(x) \cr &\ & \times \;
 \bigg[ \int_{\mathcal{M}^4} \mathrm{d}_g^2 x_1 \, \mathrm{d}_g^2 x_2 \,
\mathrm{d}_g^2 x'_1 \, \mathrm{d}_g^2 x'_2 \;
\phi^{\dag}_{g}(x'_1)\,
 \phi^{\dag}_{g}(x'_2) \int_{0}^{\infty} \mathrm{d} t\; K_{t}(x_1,x;g) \,
 K_{t}(x_2, x;g) \cr &\ & \times \; K_{t}(x',x'_1;g) \,
 K_{t}(x',x'_2;g)\, e^{-t (H_0 -E)}
\phi_{g}(x_1)\, \phi_{g}(x_2) + 4 \int_{\mathcal{M}^2}
\mathrm{d}_g^2 x_1 \, \mathrm{d}_g^2 x_2 \; \phi^{\dag}_{g}(x_1) \
\cr &\ & \times \; \int_{0}^{\infty} \mathrm{d} t \;K_{t}(x_2,x;g)
\, K_{t}(x',x;g)\, K_{t}(x',x_1;g)\, e^{-t (H_0 -E)} \phi_{g}(x_2)
 \bigg] \chi_{g}(x') \;.
\eeqs
\end{widetext}
This is exactly equal to $\Phi^{R}(\gamma^{-1}M,\lambda_R(M),E;g)$
and this is indeed $U^{\dagger}(\gamma) \Phi^{R}(M, \lambda_R(M),
\gamma^2 E;\gamma^{-2}g) U(\gamma)$ due to (\ref{transformedPhi}).
This shows that one can alternatively find out evolution of the
coupling constant which is given (\ref{coupling const evolve 2d})
from the scaling relation (\ref{grm scaling 2d}).

\section{Conclusion}

In this paper, we have constructed a new non-perturbative
renormalization method to the many-body problem on two dimensional
manifolds. The ground state energy is studied in the mean field
approximation. The renormalization group equation has been derived
and the $\beta$ function is exactly given, as a result it is shown
that the model is asymptotically free.

\section{Acknowledgments}

The authors gratefully acknowledge the many helpful
discussions with \c{C}. Dogan, B. Kaynak. We also would like to
express our deep and sincere gratitude to S. G. Rajeev for his
inspiring work. O. T. T would like to thank J. Hoppe, for his
interest in this problem and his constant support. Finally, we thank the anonymous referees for their suggestions to improve our paper.

\end{document}